\documentclass[onecolumn,amssymb,amssymb,aps,pra]{revtex4}
\usepackage{epsfig}

\begin{document}

\title{Solving procedure for a twenty-five diagonal coefficient matrix:
direct numerical solutions of the three dimensional linear
Fokker-Planck equation}

\author{Maximiliano Ujevic}\email[e-mail: ]{mujevic@ime.unicamp.br}

\author{Patricio S. Letelier}\email[e-mail: ]{letelier@ime.unicamp.br}
\affiliation{Departamento de Matem\'atica Aplicada, Instituto de 
Matem\'atica, Estat\'{\i}stica e Computa\c{c}\~ao Cient\'{\i}fica \\ 
Universidade Estadual de Campinas, 13081-970, Campinas, SP, Brasil}

\begin{abstract}

We describe an implicit procedure for solving linear equation systems
resulting from the discretization of the three dimensional (seven
variables) linear Fokker-Planck equation. The discretization of the
Fokker-Planck equation is performed using a twenty-five point molecule
that leads to a coefficient matrix with equal number of diagonals. The
method is an extension of Stone's implicit procedure, includes a vast
class of collision terms and can be applied to stationary or non
stationary problems with different discretizations in time. Test
calculations and comparisons with other methods are presented in two
stationary examples, including an astrophysical application for the
Miyamoto-Nagai disk potential for a typical galaxy.

\end{abstract}

\maketitle

\section{Introduction}

In dealing with solutions of partial differential equations we often
encounter a set of linear equations that has to be solved. This set of
linear equations depends on the method used for discretization. In
general, when dealing with three dimensional systems the number of
linear equations increases and the numerical solution of these
equations uses most of the computing time. An extreme case is the
Fokker-Planck equation. The Fokker-Planck equation is also known as the
Fokker-Planck approximation because truncates the BBGKY (N.N.
Bogoliubov, M. Born, H.S. Green, J.G. Kirkwood, and J. Yvon) hierarchy
of kinetic equations at its lowest order by assuming that correlation
between particles only plays a role as a sequence of uncorrelated
two-body encounters \cite{rei,hua}. Note that the only
``approximation'' made in the Fokker-Planck equation comes from the
model adopted for collisions and, in fact, the Fokker-Planck equation
can be derived from first principles and no {\it ad hoc} suppositions
are needed. The solution of the Fokker-Planck equation equation is not
an easy task because in the three dimensional case it has seven
variables: three space coordinates ({\bf x}), three velocity
coordinates ({\bf v}) and time ($t$). In the two dimensional case there
is a simplification because the total number of variables are five. In
either case, the large number of grid nodes needed for the computation
of the solution becomes a storage data problem. In a three dimensional
problem, for a stationary or non stationary equation, the number of
linear equations corresponds to the number of nodes in the phase-space
grid ({\bf x,v}). If we divide each of the phase-space variables'
interval of the distribution function in nine parts (ten nodes), we
will have a grid with $10^6$ nodes. In a simple numerical method we
have to store and solve a matrix with $10^{12}$ elements. For the two
dimensional case, the main matrix will have $10^8$ elements. With 10
grid nodes per variable only very simple geometries can be described.
The large number of matrix elements brings us another computational
problem, the slowness of the codes. In the discretization process of
the Fokker-Planck equation, a system of linear equations is obtained
and arranged into a matrix form (coefficient matrix). For the case of a
finite difference scheme discretization in three dimensions with a
twenty five point molecule, we see that approximately less than 0.003\%
of the elements are different from zero. This incentives us to search
for alternative and faster methods, usually iterative, to solved the
linear system using only the non null data. Note that, in general, the
coefficient matrix is not symmetric. So, powerful methods like the
Conjugate Gradient \cite{hes:sti,ger:whe} and Cholesky \cite{ger:whe}
decomposition can not be used.  Our main goal is to obtain a code that
allows us to obtain a fast and effective numerical solutions on high
resolution schemes of the three dimensional linear Fokker-Planck
equation in a direct way. The importance and difficulties of having
three dimensional solutions of the Fokker-Planck equation can be
summarized in the words of Binney and Tremaine (\cite{bin:tre} page
245), here in relation with galactic dynamics: {\it Finding the
particular function of three variables that describes any given galaxy
is no simple matter. In fact, this task has proved so daunting that
only in the last few years, three-quarters of a century after Jeans's}
(Jeans \cite{jea}) {\it paper posed the problem, has the serious quest
for the distribution function of even our own Galaxy got underway.}

We mean by direct numerical calculations of the Fokker-Planck equation
a method that is neither statistical nor mean field approximation
\cite{bin:tre}. Numerical solutions can be performed using statistical
approximate methods like the method of moment equations
\cite{lar1,lar2,lyn:egg} and Monte Carlo methods \cite{hen1,hen2}.
Also, solutions have been found using the orbit-averaged Fokker-Planck
equation with action-angle variables \cite{bin:tre}, this last method
reduces the equation involving six phase-space coordinates plus time to
one involving only three actions plus time. Another method that have
been used for solving the Vlasov equation with good results is the
operator splitting technique \cite{che:kno,gag:sho}. Two dimensional
numerical integration of the Vlasov equations can be found in
\cite{sho:ger}. The problem associated with this method is that, even
in the two dimensional case, the time spent in making the interpolation
is very expensive. Direct three dimensional numerical calculations of
the Fokker-Planck equations, to the best of our knowledge, has not been
done. As we said before, the main difficulty to find the solution of
the three dimensional Fokker-Planck equation is the large number of
linear equations that is translated in a high computational cost
involve in the process. In this article we present a variations of
Stone's \cite{sto} method that leads us to solve with low computational
cost the three dimensional linear Fokker-Planck equation. Variations of
Stone's method has been applied to other situations in two
\cite{sch:zed} and three \cite{lei:per} space dimensions when dealing
with fluid flow, heat transfer and Laplace equation.

In this article we study how to solve the system of equations that
arises from the discretization process using a finite difference
scheme, in particular we used a twenty-five point molecule, but we have
to keep in mind that other discretization procedures gives practically
the same form for the coefficient matrix and this method can also be
applied, i.e. sparse matrices with the same number of diagonals. For
example, in two dimensions, the diffusion equation can be discretized
in an uniform rectangular grid by using the finite difference scheme,
the finite volume method (which is a discretization of the equation in
integral form) or the finite element method, see \cite{fle}. All of
these methods gives five diagonals, in the finite element method the
number of diagonals depends on the type of interpolation considered,
and the coefficient matrix can be solved by the same numerical method.
So, in this article, we shall not discuss what discretization method
is better for the problem considered, we rather present a numerical
method that allows us to find the solution of a coefficient matrix with
twenty-five diagonals that can be obtained with different
discretization methods.

The article is organized as follows. In Section II we present the
linear Fokker-Planck equation to be solved. We consider a linear
collision term that includes a vast class of collisions. In Section
III, we describe the algorithm of the modified Stone method to obtain
the incomplete $LU$ decomposition. The discretization of the
Fokker-Planck equation is performed using a central difference
approximation for the phase-space variables ({\bf x,v}) that is
described with a twenty-five point molecule. This point molecule also
allows to describe different discretizations in time, as the implicit
Euler or Crank-Nicolson discretization. The derivation of the $LU$
decomposition is made without assuming any particular discretization in
time, maintaining its derivation as general as possible. The notation
used for the diagonals in the coefficient matrix is also shown. In
Section IV, we test our algorithm with an astrophysical example by
solving the Fokker-Planck equation for the widely used Miyamoto-Nagai
disk potential \cite{miy:nag} in three dimensions. The parameters of
the model are chosen to represent the Newtonian potential of a typical
galaxy. Also, we used a Fokker-Planck test equation to compare our
algorithm with the Generalized Minimal Residual Method (GMRES) that
solves large sparse matrices. In Section V, we show how to modified the
code to implement curved boundary conditions. Finally, in Section VI,
we summarized our results.

\section{The general problem}

The linear Fokker-Planck equation can be written as

\begin{equation}
\frac{\partial f}{\partial t} + {\bf v} \cdot \nabla f + \dot{\bf v}
\cdot  \nabla_v f = \Gamma[f],
\label{fokker}
\end{equation}

\noindent where ${\bf v}$ represent the velocity of the particles,
$\nabla$ is the usual gradient, $\nabla_v$ is the velocity gradient
(derivations are done with respect to the velocities), $\dot{\bf v}$ is
the acceleration and the symbol $\Gamma[f]$ denotes the rate of change
of $f$ due to encounters (collision term). We consider as the collision
term the expression

\begin{equation}
\Gamma[f] = A({\bf x,v})  \nabla^2 f + B({\bf x,v}) \nabla^2_v f + 
C({\bf x,v}) \nabla f + D({\bf x,v}) \nabla_v f + \sum_{i \neq j=1}^3 
E_{ij}({\bf x,v}) \frac{\partial^2 f}{\partial v_i \partial v_j}, 
\label{colterm}
\end{equation}

\noindent where $A$, $B$, $C$, $D$ and $E_{ij}$ are arbitrary functions
of the phase-space variables ${\bf x}$ and ${\bf v}$. The equation
above describes a vast family of collisions. In particular, with the
mixed velocity derivative term present in (\ref{colterm}) we can take
into account the important collision term found by Rosenbluth et al.
\cite{ros:mac} used in gravitating systems and plasma physics, see for
example \cite{ein:spu,gal:mac} and \cite{nan,bro:pre} respectively, and
reference therein. If we need mixed space derivatives instead of mixed
velocity derivatives, we can use the same code presented in this
article to solve the problem. If a particular problem requires the
inclusion of mixed velocity derivatives as well as mixed space
derivatives, it is possible to develop a similar numerical procedure
following the steps of this article, but it complicates the incomplete
$LU$ decomposition used for the method, i.e. we need twelve extra
diagonals on the coefficient matrix (thirty-seven instead of
twenty-five diagonals). 

In Section IV we solve numerically the stationary linear Fokker-Planck
equation (\ref{fokker}). In general, Eq. (\ref{fokker}) is non-linear
because $\dot{\bf v} = {\bf F}(f)/m$, where ${\bf F}$ is the force and
$m$ is the mass of the particle. Another kind of non-linearity may
arrise from the collision term considered. A way to deal with this kind
of problem is to start with a given distribution function at time $t$
from which we can calculate the force and collision term at this time.
Then, this force and collision term are replaced into the Fokker-Planck
equation from which we obtain the distribution function for a later
time $t+\Delta t$. With the recently calculated distribution function
we can calculate again the force and collision term at time $t + \Delta
t$ and the process is repeated. An application for a stationary
non-linear problem can be found in \cite{uje:let}, in which we found
the distribution function that satisfies both Fokker-Planck and Poisson
equation in two dimensions for a Kuzmin-Toomre thin disk.

\begin{table}
\caption{\label{discret} Nomenclature and relations between the matrix 
form and the one dimensional storage index at node $p$ of the 
twenty-five terms used in the discretization of the Fokker-Planck 
equation.}

\begin{tabular}{|cccc|}
\hline
Matrix Form & Abbreviation & Name & Position From Node $p$ \\
\hline
& \multicolumn{2}{c}{\rm \bf Basic Diagonals} &  \\
$\Psi(i,j,k,l,m,n)$ & $P$ & point & $p$ \\
$\Psi(i,j+1,k,l,m,n)$ & $N$ & north & $p+1$ \\
$\Psi(i,j-1,k,l,m,n)$ & $S$ & south & $p-1$ \\
$\Psi(i+1,j,k,l,m,n)$ & $E$ & east & $p+n_j$ \\
$\Psi(i-1,j,k,l,m,n)$ & $W$ & west & $p-n_j$ \\
$\Psi(i,j,k+1,l,m,n)$ & $T$ & top & $p+n_{ij}$ \\
$\Psi(i,j,k-1,l,m,n)$ & $B$ & bottom & $p-n_{ij}$ \\
$\Psi(i,j,k,l+1,m,n)$ & $U1$ & up1 & $p+n_{ijk}$ \\
$\Psi(i,j,k,l-1,m,n)$ & $D1$ & down1 & $p-n_{ijk}$ \\
$\Psi(i,j,k,l,m+1,n)$ & $U2$ & up2 & $p+n_{ijkl}$ \\
$\Psi(i,j,k,l,m-1,n)$ & $D2$ & down2 & $p-n_{ijkl}$ \\
$\Psi(i,j,k,l,m,n+1)$ & $U3$ & up3 & $p+n_{ijklm}$ \\
$\Psi(i,j,k,l,m,n-1)$ & $D3$ & down3 & $p-n_{ijklm}$ \\
& \multicolumn{2}{c}{\rm \bf Mixed Diagonals} &  \\
$\Psi(i,j,k,l+1,m+1,n)$ & $UT2$ & uptop2 & $p+(n_{ijkl} + n_{ijk})$ \\
$\Psi(i,j,k,l-1,m+1,n)$ & $UB2$ & upbottom2 & $p+(n_{ijkl} - n_{ijk})$ \\
$\Psi(i,j,k,l+1,m-1,n)$ & $DT2$ & downtop2 & $p-(n_{ijkl} - n_{ijk})$ \\
$\Psi(i,j,k,l-1,m-1,n)$ & $DB2$ & downbottom2 & $p-(n_{ijkl}+n_{ijk})$ \\
$\Psi(i,j,k,l+1,m,n+1)$ & $UT3$ & uptop3 & $p+(n_{ijklm} + n_{ijk})$ \\
$\Psi(i,j,k,l-1,m,n+1)$ & $UB3$ & upbottom3 & $p+(n_{ijklm} - n_{ijk})$ \\
$\Psi(i,j,k,l+1,m,n-1)$ & $DT3$ & downtop3 & $p-(n_{ijklm} - n_{ijk})$ \\
$\Psi(i,j,k,l-1,m,n-1)$ & $DB3$ & downbottom3 & $p-(n_{ijklm}+n_{ijk})$ \\
$\Psi(i,j,k,l,m+1,n+1)$ & $UU3$ & upup3 & $p+(n_{ijklm}+n_{ijkl})$ \\
$\Psi(i,j,k,l,m-1,n+1)$ & $UD3$ & updown3 & $p+(n_{ijklm}-n_{ijkl})$ \\
$\Psi(i,j,k,l,m+1,n-1)$ & $DU3$ & downup3 & $p-(n_{ijklm}-n_{ijkl})$ \\
$\Psi(i,j,k,l,m-1,n-1)$ & $DD3$ & downdown3 & $p-(n_{ijklm}+n_{ijkl})$ \\
\hline
\end{tabular}
\end{table}

\section{Description of the algorithm} \label{description}

The system of equations obtained from the discretization of the
Fokker-Planck equation (\ref{fokker},\ref{colterm}) using the central
finite difference approximation for the phase-space and a temporal
discretization in time (implicit Euler, Crank-Nicolson, etc.) can be
cast (for each  time step) into the simple form,

\begin{equation}
A \Psi = Q, \label{soe}
\end{equation}

\noindent in which $A$ is a square coefficient matrix $N_{node} \times
N_{node}$ ($N_{node}$ the number of nodes in the discretization grid),
$\Psi$ is the vector matrix of the nodal variable values, and $Q$ is
the source vector. The position of the grid nodes in the phase-space
($x,y,z,v_x,v_y,v_z$) is performed by six indexes ($i,j,k,l,m,n$),
where $i$ represents the index for the variable $x$, $j$ represents the
index for the variable $y$, etc. The ordering of nodes in this six
dimensional space is made as follows. The surface $n$=constant are
stacked one above another. Within the fifth dimensional space (for each
$n$) the hyper-surfaces $m$=constant are stacked one above another.
Within the fourth dimensional space (for each $n$ and $m$) the
hyper-surfaces $l$=constant are stacked one above another. Within the
three dimensional space (for each $n$, $m$ and $l$) the surfaces
$k$=constant are stacked one above another. Within the two dimensional
space (for each $n$, $m$, $l$ and $k$) the index $j$ increases first
($y$-direction) than the index $i$ ($x$-direction). The one dimensional
storage index $p$ of the vector matrix $\Psi$ is calculated from the
six-dimensional grid indexes ($i,j,k,l,m,n$), i.e.

\begin{equation}
p=(n-1) n_{ijklm} + (m-1) n_{ijkl} + (l-1) n_{ijk} +(k-1) n_{ij} +(i-1) 
n_j+ j, \label{p}
\end{equation}

\noindent with

\begin{eqnarray}
&&i=1 \cdots n_i; \;\; j=1 \cdots n_j; \;\; k=1 \cdots n_k; \;\; l=1 
\cdots n_l; \;\; m=1 \cdots n_m; \;\; n=1 \cdots n_n; \nonumber \\
&&n_{ijklm} = n_i n_j n_k n_l n_m; \;\; n_{ijkl} = n_i n_j n_k n_l; \;\;  
n_{ijk}=n_i n_j n_k; \;\; n_{ij}= n_i n_j;
\end{eqnarray}

\begin{figure}
\begin{center}
\epsfig{width=12cm,height=12cm, file=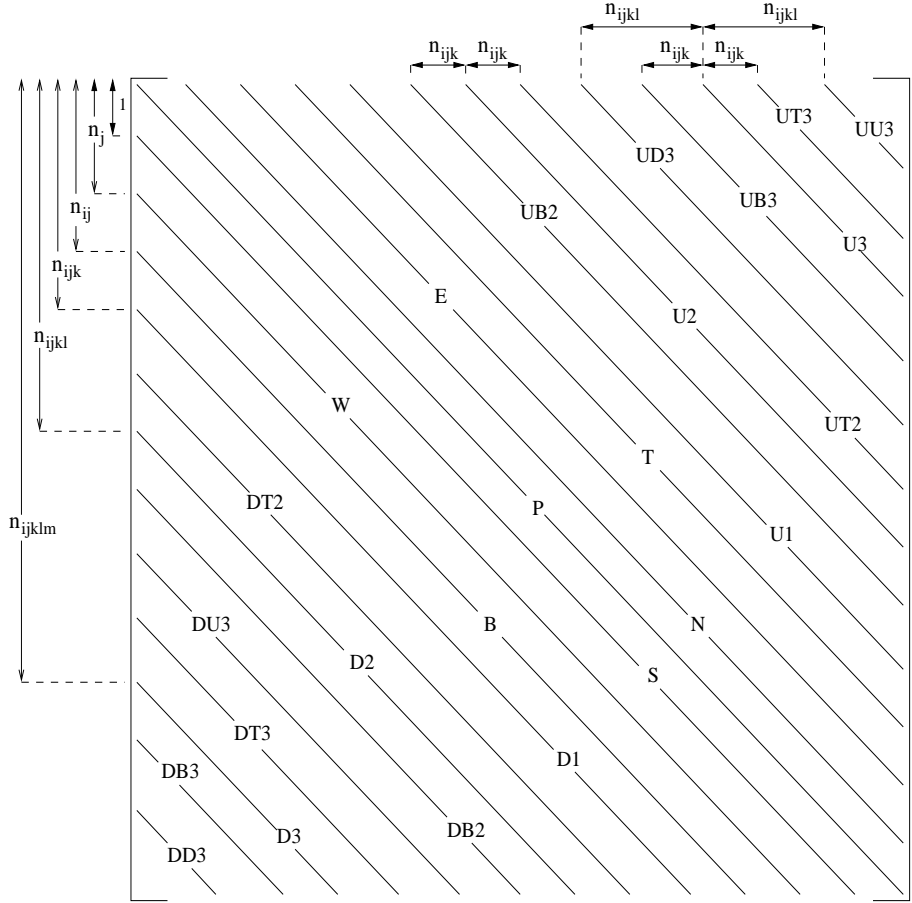}
\end{center}
\caption{Form of the coefficient matrix $A$ obtained from the
discretization of the partial difference equation of our problem. The
separation between the main diagonal $P$ and the other diagonals are
indicated. Also, the distance between the diagonals of the mixed
derivatives to their nearest diagonals are shown. Note that the figure is
not in scaled.} \label{figdiagonals}
\end{figure}

\begin{figure}
\epsfig{width=17cm,height=5cm, file=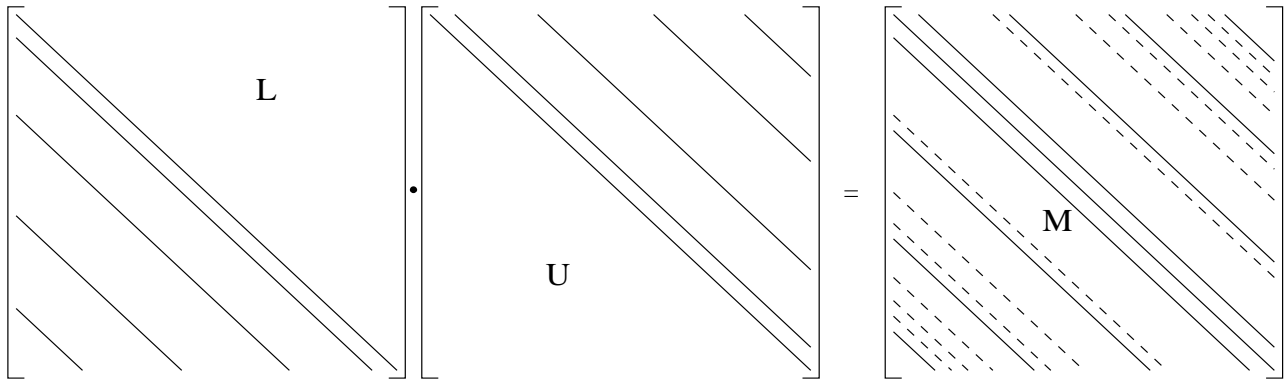}
\caption{Schematic representation of the matrices $L$, $U$ and their
product $M$. The multiplication of the matrices $L$ and $U$ leads to extra
diagonals (dotted lines) not present in the coefficient matrix $A$.  The
diagonals of $L$, from the left-bottom corner to the main diagonal, are:  
$DD3$, $DB3$, $D3$, $DT3$, $DU3$, $DB2$, $D2$, $DT2$, $D1$, $W$, $S$, $P$.  
The diagonals of $U$, from the main diagonal to the right-up diagonal,
are: $1$, $N$, $E$, $T$, $U1$, $UB2$, $U2$, $UT2$, $UD3$, $UB3$, $U3$,
$UT3$, $UU3$. The diagonals in $M$ are products of these two sets of 
diagonals but we have to be careful because more than one product can be 
at the same diagonal.}
\label{matrixm}
\end{figure}

\noindent where $n_i$, $n_j$, $n_k$, $n_l$, $n_m$, $n_n$ denote the
number of grid points for each variable. Therefore $N_{node} = n_i n_j
n_k n_l n_m n_n$. With the help of the storage index $p$ we can switch
each point of the twentyfive-point molecule from the matrix form to the
one dimensional position representation. This is done in Table
\ref{discret} by making the equivalence $f(x,y,z,v_x,v_y,v_z) \equiv
\Psi(i,j,k,l,m,n)$. Until now, we considered only the non stationary
case, but the discretization in Table \ref{discret} can be used in the
non stationary as well as the stationary Fokker-Planck equation. In
both cases, the final system of equations is of the form (\ref{soe}),
with $A$ being a sparse matrix with elements different from zero in
only twenty-five diagonals, see Fig. \ref{figdiagonals}. Now, we want
to develop an iteration method in order to solve the system of
equations (\ref{soe}). After $h$ iterations of such method, the
approximate solution $\Psi^{h}$ do not satisfies (\ref{soe}) exactly,
their is a non zero residual $R$ such as

\begin{equation}
A \Psi^{h} = Q - R^{h}.
\end{equation}

\noindent The purpose of the iteration procedure is to drive the
residual term to zero after some number of iteration (actually we stop
the iteration when the residual term attained some imposed small value
condition). Let us consider an iterative scheme for a linear system in
the form

\begin{equation}
M \Psi^{h+1} = O \Psi^h + B, \label{itproce}
\end{equation}

\noindent when convergence is achieved we must have that $A=M - O$ and
$B=Q$. An alternative version of this procedure can be obtained by
subtracting $M \Psi^{h}$ from both sides of (\ref{itproce}) to have,

\begin{equation}
M \Delta^{h+1} = R^{h}, \label{iteration}
\end{equation}

\noindent where $\Delta^{h+1} = \Psi^{h+1} - \Psi^{h}$ and $R^{h} =
B-(M - O) \Psi^{h} = Q - A \Psi^h$. Any effective iterative method to
solve (\ref{itproce}) must be cheap and converge rapidly. For faster
convergence, the matrix $M$ have to be a good approximation of the
coefficient matrix $A$, i.e. we must have $O \Psi^h$ small. The
original idea of Stone is to use for the iteration matrix $M$ an
incomplete $LU$ decomposition of the matrix $A$. The reason for this
choice is that $LU$ decomposition is an excellent linear system solver.
The matrices $L$ and $U$ have elements different from zero only in the
diagonals in which $A$ have also elements different from zero. The
product of the matrices, $L$ and $U$, provide a matrix $M$ with a
larger number of diagonals with elements different from zero, see for
instance Fig. \ref{matrixm}. To make the decomposition $LU$ unique, we
set the elements on the principal diagonal of $U$ equal to 1. In doing
the $LU$ multiplication, we have to pay extra attention because
sometimes more than one diagonal product are at the same distance from
the main diagonal $P$, e.g. the product $L_{D3} \cdot U_1$ and $L_{DD3}
\cdot U_{U2}$ are at the same diagonal in $M$. The multiplication rules
for matrices furnish the elements of $M=LU$ at node $p$ (see Appendix
\ref{MLU}), where for example, $M_{W|UU3}$ represents the diagonal in
$M$ that is obtain from the multiplication between the elements of the
diagonal $S$ in $L$ with the elements of the diagonal $UU3$ in $U$.

Now, we choose $L$ and $U$ in such a way that $M(=A+O)$ is the best
possible approximation to $A$. The standard method for decomposition is
to let $O$ to have elements different from zero in the diagonals of $M$
that corresponds to the diagonals not present in $A$ and to force the
other diagonals of $M$ to be equal to the corresponding diagonals in
$A$.  But this method converges slowly because the resulting matrix $O$
is not small. Stone recognized that the convergence of the method could
be faster if we allow $O$ to have elements different from zero also in
the diagonals present in $A$. The key idea is that the contribution of
$M \Psi$ of the diagonals not present in $A$ partially canceled the
contribution of $O\Psi$ of the diagonals present in $A$, in such a way
that

\begin{equation}
O \Psi \approx 0. \label{Oapprox}
\end{equation}

\noindent Note that in (\ref{relaM}) they are diagonals in $M$ that
present more than one term. In general, the principal diagonals have
more than one element, as for example $M_{DT3}$, but beside these
principal diagonals there are other no-principal diagonals that have
more than one element, like $M_{D3|UB2}$. Now, relation (\ref{Oapprox})
can be written for one grid node in several ways.  The usual way is to
consider the elements of these no-principle diagonals as part of the
same diagonal. Other way is to consider these elements as they were
from different diagonals, thus in this case, following the above
example, the no-principal diagonal $M_{D3|UB2}$ is split into two
diagonals $M_{D3|UB2}$ and $M_{DB3|U2}$. We obtained the final
relations for the $LU$ decomposition in both ways and we found that the
$LU$ decomposition considering the elements as they were from different
diagonals is faster by a factor of two. Hereafter we consider this
case. The explicit form of equation (\ref{Oapprox}) is given in
Appendix \ref{Oexplicit}.

The problem now is to defined the elements of $O$ to satisfy Eq.
(\ref{noderelation}) without introducing additional unknowns. If we
expect the solution of the partial differential equation to be smooth,
we can approximate the values of $\Psi_{B|N}$, $\Psi_{N|W}$, etc, in
terms of the values of $\Psi$ at nodes corresponding to the diagonals
of A. Stone proposed the following approximations (other approximations
are possible),

\begin{eqnarray}
&&\Psi_{B|N} \approx \alpha (\Psi_B + \Psi_N - \Psi_P), \nonumber \\
&&\Psi_{W|N} \approx \alpha (\Psi_W + \Psi_N - \Psi_P), \; {\rm etc}
\label{combination}
\end{eqnarray}

\noindent where $\alpha$ is a constant. Stability analysis made by
Stone requires that $\alpha$ must be between $ 0 < \alpha < 1$.
Replacing the above approximations into (\ref{noderelation})  we obtain
the elements of $O$ as a linear combination of the elements of $M$, see
for instance Appendix \ref{linear}. Now, using the relation $M=A+O$
together with expressions (\ref{relaM})  and (\ref{relaO}), we find
that the elements of the matrices $L$ and $U$ are given by,

\begin{eqnarray}
&&L_{[X]}^p = \frac{A_{[X]}^p - C_{[X]}}{1 + \alpha K_{[X]}}, \;\;\; [X] = 
DD3,\;DB3,\;D3,\;DT3,\;DU3,\;DB2,\;D2,\;DT2,\;D1,\;B,\;W,\;S, \nonumber \\
&&L_{P}^p = A_{P}^p - L_{S}^p U_{N}^{p -1} - L_{W}^p U_{E}^{p -n_j} -
L_{B}^p U_{T}^{p - n_{ij}} - L_{D1}^p U_{U1}^{p - n_{ijk}} - L_{DT2}^p
U_{UB2}^{p -(n_{ijkl} - n_{ijk})} - L_{D2}^p U_{U2}^{p - n_{ijkl}} 
\nonumber \\
&&\hspace{1cm}- L_{DB2}^p U_{UT2}^{p - (n_{ijkl} + n_{ijk})} - L_{DU3}^p 
U_{UD3}^{p - (n_{ijklm} - n_{ijkl})} - L_{DT3}^p U_{UB3}^{p - (n_{ijklm} - 
n_{ijk})} - L_{D3}^p U_{U3}^{p -n_{ijklm}} \nonumber \\
&&\hspace{1cm}- L_{DB3}^p U_{UT3}^{p - (n_{ijklm} + n_{ijk})} - L_{DD3}^p 
U_{UU3}^{p - (n_{ijklm} + n_{ijkl})} + \alpha (K_{N} + K_{E} + K_{T} + 
K_{U1} + K_{UB2} + K_{U2} + K_{UT2} \nonumber \\
&&\hspace{1cm} + K_{UD3} + K_{UB3} + K_{U3} + K_{UT3} + K_{UU3}), 
\nonumber \\
&&U_{[X]}^p = \frac{A_{[X]}^p - \alpha K_{[X]} - C_{[X]}}{L_{P}^p}, \;\;\;  
[X] = N,\;E,\;T,\;U1,\;UB2,\;U2,\;UT2,\;UD3,\;UB3,\;U3,\;UT3,\;UU3, 
\label{LUdecom}
\end{eqnarray}

\noindent where the explicit form of the functions $C_{[X]}$ and
$K_{[X]}$ are given in the Appendix \ref{CandK}. The elements of the
$LU$ decomposition have to be calculated in the order specified in
(\ref{LUdecom}). In doing this, we must take into account that a
certain element is considered equal to zero if its storage index is
less or equal zero, e.g. if $p=3$ and $n_j = 5$ then the elements with
index $p$ and $p-1$ are different from zero, and the elements with
index $p - n_j$, $p - n_{ij}$, etc are equal to zero. When mixed
derivatives are not present we must set all the elements with index
$DD3$, $DB3$, $DT3$, $DU3$, $DB2$, $DT2$, $UB2$, $UT2$, $UD3$, $UB3$,
$UT3$, $UU3$ equal to zero. Once obtained the $LU$ decomposition, the
system of equation is solved combining $M=LU$ with (\ref{iteration}) to
obtain

\begin{equation}
LU \Delta^{h+1} = R^{h},
\end{equation}

\noindent and here we set

\begin{eqnarray}
&&L \Upsilon^{h} = R^{h}, \nonumber \\
&&U \Delta^{h+1} = \Upsilon^{h}, 
\end{eqnarray}

\noindent from which we obtain the solution of our problem by solving
two triangular systems. In this iterative method, the matrix elements
of $L$ and $U$ are calculated only once before the first iteration. In
other iterations, only the residual $R$, $\Upsilon$ and $\Delta$ are
calculated using the two triangular system mentioned above, i.e.,

\begin{eqnarray}
&&\Upsilon^p = (R^p - L^p_{DD3} \Upsilon^{p- (n_{ijklm} + n_{ijkl})} -
L^p_{DB3} \Upsilon^{p- (n_{ijklm} + n_{ijk})} - L^p_{D3} \Upsilon^{p-
n_{ijklm}} - L^p_{DT3} \Upsilon^{p- (n_{ijklm} - n_{ijk})} \nonumber \\
&&\hspace{1cm}- L^p_{DU3} \Upsilon^{p- (n_{ijklm} - n_{ijkl})} - L^p_{DB2} 
\Upsilon^{p- (n_{ijkl} + n_{ijk})} - L^p_{D2} \Upsilon^{p- n_{ijkl}} - 
L^p_{DT2} \Upsilon^{p- (n_{ijkl} - n_{ijk})} - L^p_{D1} \Upsilon^{p- 
n_{ijk}} \nonumber \\
&&\hspace{1cm}- L^p_{B} \Upsilon^{p- n_{ij}} - L^p_{W} \Upsilon^{p- n_j} - 
L^p_{S} \Upsilon^{p- 1})/L^p_{P}, \nonumber \\
&&\Delta^p = \Upsilon^p - U^p_{UU3} \Delta^{p+( n_{ijklm} + n_{ijkl})} -
U^p_{UT3} \Delta^{p+( n_{ijklm} + n_{ijk})} - U^p_{U3} \Delta^{p+
n_{ijklm}} - U^p_{UB3} \Delta^{p+( n_{ijklm} - n_{ijk})} \nonumber \\
&&\hspace{1cm}- U^p_{UD3} \Delta^{p+( n_{ijklm} - n_{ijkl})} - U^p_{UT2} 
\Delta^{p+( n_{ijkl} + n_{ijk})} - U^p_{U2} \Delta^{p+ n_{ijkl}} - 
U^p_{UB2} \Delta^{p+( n_{ijkl} - n_{ijk})} - U^p_{U1} \Delta^{p+ n_{ijk}} 
\nonumber \\
&&\hspace{1cm}- U^p_{T} \Delta^{p+ n_{ij}} - U^p_{E} \Delta^{p+ n_{j}} - 
U^p_{N} \Delta^{p+ 1},
\end{eqnarray}

\noindent where

\begin{eqnarray}
&&R^p = Q^p - A_{P}^p \Psi^p - A_{N}^p \Psi^{p+1} - A_{S}^p \Psi^{p-1} -
A_{E}^p \Psi^{p+n_j} - A_{W}^p \Psi^{p-n_j} - A_{T}^p \Psi^{p+n_{ij}} -
A_{B}^p \Psi^{p-n_{ij}} - A_{U1}^p \Psi^{p+n_{ijk}} \nonumber \\
&&\hspace{1cm} - A_{D1}^p \Psi^{p-n_{ijk}} - A_{UB2}^p 
\Psi^{p+(n_{ijkl}-n_{ijk})} - A_{DT2}^p \Psi^{p-(n_{ijkl}-n_{ijk})} - 
A_{U2}^p \Psi^{p+n_{ijkl}} - A_{D2}^p \Psi^{p-n_{ijkl}} \nonumber \\
&&\hspace{1cm} - A_{UT2}^p \Psi^{p+(n_{ijkl}+n_{ijk})} - A_{DB2}^p 
\Psi^{p-(n_{ijkl}+n_{ijk})} - A_{UD3}^p \Psi^{p+(n_{ijklm}-n_{ijkl})} 
- A_{DU3}^p \Psi^{p-(n_{ijklm}-n_{ijkl})} \nonumber \\
&&\hspace{1cm}- A_{UB3}^p \Psi^{p+(n_{ijklm}-n_{ijk})} - A_{DT3}^p 
\Psi^{p-(n_{ijklm}-n_{ijk})} - A_{U3}^p \Psi^{p+n_{ijklm}} - 
A_{D3}^p \Psi^{p-n_{ijklm}} \nonumber \\
&&\hspace{1cm}- A_{UT3}^p \Psi^{p+(n_{ijklm}+n_{ijk})} - A_{DB3}^p 
\Psi^{p-(n_{ijklm}+n_{ijk})} - A_{UU3}^p \Psi^{p+(n_{ijklm}+n_{ijkl})} 
- A_{DD3}^p \Psi^{p-(n_{ijklm}+n_{ijkl})}, \nonumber
\end{eqnarray}

\noindent in which, for simplicity, we have omitted the iterative index
$h$.

\section{Test results and comparisons with other methods}

As was mentioned in the Introduction, the huge number of nodes needed
to solve the Fokker-Planck equation leads to large amount of data that
has to be stored in a matrix, this fact reduces the possible codes for
testing the results. Furthermore, in general, the Fokker-Planck
equation (\ref{fokker}) is not symmetric, and for that reason other
methods can not be used. In this case, the Generalized Minimal Residual
method (GMRES) \cite{saa:sch} is the most appropriate choice. The GMRES
method belongs to the class of Krylov based iterative methods
\cite{lan1,arn,lan2} and was proposed in order to solve large, sparse
and non Hermitian linear systems. This section is divided into two
parts. In the first part, we apply the method presented in this article
to a physical example. In particular, we choose the astrophysical
problem of finding the distribution function of the Miyamoto-Nagai disk
in three dimensions. In the second part, we used a simplified version
of the physical example to make numerical comparisons. We compare our
method with the best available method in solving huge sparse matrices,
i.e. GMRES. We prefer to make the numerical comparisons with the
simplified version of the physical example because the large amount of
parameters present in the physical example make the analysis
cumbersome. This analysis is important because we show some advantages
and limitations of our method.

\subsection{Physical Example}
 
To test of our code we begin with an important astrophysical
application, i.e. solve the Fokker-Planck equation to find the
distribution function of the Miyamoto-Nagai disk \cite{miy:nag} in
three dimensions in a stationary regime. The Fokker Planck equation to
be solved is \cite{bin:tre}

\begin{equation}
{\bf v} \cdot \nabla f + \dot{\bf v} \cdot \nabla_v f =  - \sum_{i=1}^3
\frac{\partial}{\partial v_i} [ f({\bf x,v}) D(\Delta v_i)] +
\frac{1}{2} \sum_{i,j=1}^3 \frac{\partial^2}{\partial v_i \partial v_j}
 [ f( {\bf x,y}) D (\Delta v_i \Delta v_j) ], 
\label{testMN}
\end{equation}

\noindent where $\dot{\bf v}= - \nabla \Phi$, $\Phi = - \frac{G
M}{\sqrt{x^2+y^2+(a+\sqrt{z^2 + b^2})^2}}$, $G$ is the gravitational
constant, $M$ is the total mass of the system, $a$ and $b$ are
parameters that depending on the choice the potential can represent
anything from an infinitesimal thin disk to a spherical system, and the
functions $D(\Delta v_i)$ and $D(\Delta v_i \Delta v_j)$ are known as
the diffusion coefficients. These diffusion coefficients were
calculated by Rosenbluth et al. \cite{ros:mac} considering a test star
of mass $m$ moving through an infinite homogeneous sea of field stars
of mass $m_a$ who has mean velocity equal to zero. Moreover, the
interaction between the particles are ruled by an inverse square force,
and also, each stellar encounter involve only a single pair of stars
and are independent of all others. These diffusion coefficients are
simplified if the field stars distribution function is a Maxwellian
distribution. The explicit form of these coefficients are
\cite{bin:tre}.

\begin{eqnarray}
&&D(\Delta v_i) = \frac{v_i}{v} D(\Delta v_{||}), \\
&&D(\Delta v_i \Delta v_j) = \frac{v_i v_j}{v^2} [ D(\Delta v^2_{||}) -
\frac{1}{2} D (\Delta v^2_{\bot}) ] + \frac{1}{2} \delta_{ij} D (\Delta
v^2_{\bot}), 
\end{eqnarray}

\noindent where $D(\Delta v_{||})$, $D(\Delta v^2_{||})$ and
$\Delta v^2_{\bot}$ are given by

\begin{eqnarray} 
&&D (\Delta v_{||}) = - \frac{4 \pi G^2 \rho (m + m_a) \ln \Lambda {\rm
G}(X)}{\sigma^2}, \\
&&D (\Delta v^2_{||}) = \frac{4 \sqrt{2} \pi G^2 \rho m_a \ln
\Lambda}{\sigma} \frac{{\rm G}(X)}{X}, \\
&&D (\Delta v^2_{\bot}) = \frac{4 \sqrt{2} \pi G^2 \rho m_a \ln
\Lambda}{\sigma} \left[ \frac{{\rm erf}(X) - {\rm G}(X)}{X} \right],
\end{eqnarray}

\begin{figure} 
\epsfig{width=10cm,height=7cm,file=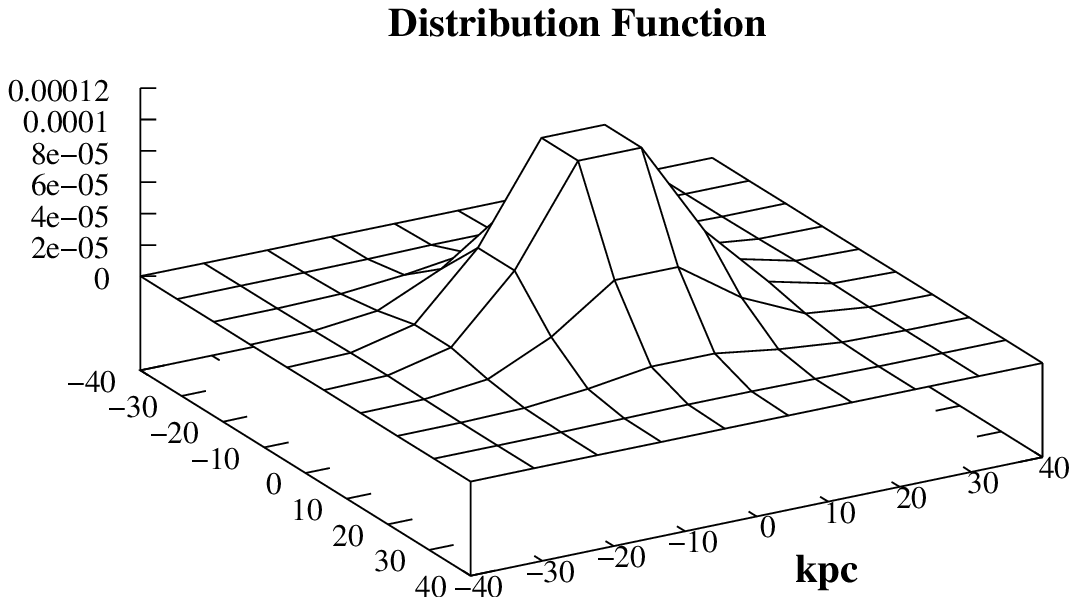}

\vspace{1cm} 
\epsfig{width=10cm,height=7cm,file=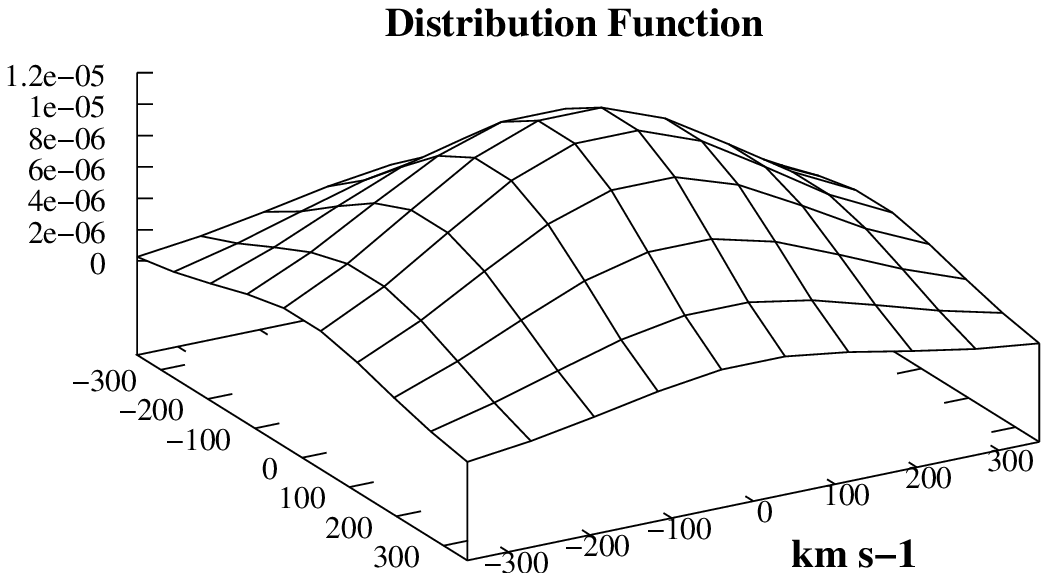}
\caption{Numerical solutions of the distribution function of Eq. 
(\ref{testMN}). Top: surface distribution for the case with $v_{x} 
\approx -117$ km s$^{-1}$, $v_y \approx 39$ km s$^{-1}$, $v_z \approx
-272$ km s$^{-1}$ and $z \approx 0.55$ kpc. Bottom:  velocity
distribution for the case with $x \approx -22.23$ kpc, $y \approx
13.32$ kpc, $z \approx -0.11$ kpc and $v_x \approx 117$ km s$^{-1}$.} 
\label{distfigMN}
\end{figure}

\noindent where $\rho$ and $\sigma$ are the density and velocity
dispersion of the field stars respectively, $X \equiv v/(\sqrt{2}
\sigma)$, ${\rm erf}(X)$ is the error function, ${\rm G}(X)=
\frac{1}{2X^2} \left[{\rm erf}(X) - \frac{2X}{\sqrt{\pi}} e^{-X^2}
\right]$, $\Lambda = \frac{b_{max} v^2_{\rm typ}}{G(m+m_a)}$, $b_{max}$
is the maximum possible impact parameter (usually set of order the
radius of the system), and $v_{\rm typ}$ is a typical velocity of stars
in the system. We shall find the solution of Eq. (\ref{testMN}) in a
six dimensional `box' using ten grid nodes for variable that leads to a
problem involving $10^6$ unknowns. If we want to solve this set of
equations using a conventional solver we need to storage $10^{12}$
elements in the coefficient matrix. We considered that the ranges of
the velocities $v_x$, $v_y$, and $v_z$ are between [-350 km s$^{-1}$,
350 km s$^{-1}$], the ranges of the coordinates $x$ and $y$ are between
[-40 kpc, 40 kpc], and the coordinate $z$ between [-1 kpc, 1 kpc]. At
the borders of the `box' we used a Dirichlet boundary condition of the
form $f=\exp[-(x^2+y^2+z^2)/2 \sigma_s^2] \exp[-(v_x^2+v_y^2+v_z^2)/2
\sigma_v^2]$. We set the parameters $a = 4$ kpc, $b =1$ kpc, $m_a=m =
M_\odot$, $\sigma_s = 10$ kpc, $\sigma_v =100$ km s$^{-1}$, $v_{\rm
typ} = 200$ km s$^{-1}$, $b_{max} = 40$ kpc and the total mass $
M=10^{12} M_\odot$. The values set for the parameters correspond to
typical galaxy data like the Milky Way. Also, as a first approximation
we spread the total mass uniformly along the disk and set the density
$\rho=$constant. In Fig. \ref{distfigMN}, we present two graphs of the
distribution function that represent the numerical solution of
(\ref{testMN}). One is the surface distribution function on the plane
$x-y$ and the other is the velocity distribution function on the plane
$v_y-v_z$; they are plotted at different points of the grid. We note
that these figures have Maxwellian form as it should.


\subsection{Numerical Comparison}

In the next example we shall compare our code based in the $LU$
decomposition (\ref{LUdecom}) with the GMRES algorithm. The problem of
this method is that the storage of the orthonormal basis may become
prohibitive for some matrices, this storage depends on the value of the
restarting parameter. The restarting parameter of the GMRES algorithm
determine the number of the orthonormal vectors for the Krylov subspace
that the code stores in order to calculate the updated solution and
residual at each time step. At each time step the code stores one
vector. After a number of steps equal to the restarting parameter, the
code construct the most recent update and use it as a first guess to
restart the next set of iterations. The convergence of the method is
guaranteed for large numbers of the restarting parameter, but this
means that more vectors have to be stored and computer memory problems
may appear. We can use lower values of the restarting parameter but
this increases the time spent in finding the solution. In particular,
we used the {\it gmres} routine of Matlab because it can handle sparse
matrices. We also used a public GMRES software \cite{fra:gir}, this
software allows us to choose between different kinds of preconditioners
and orthogonalization procedures but its drawback is that can not
handle sparse matrices. We choose as a test equation a stationary
Fokker-Planck equation similar to the physical example of the previous
section. Note that in the non stationary Fokker-Planck equation we can
perform a Crank-Nicolson discretization in time, which is also an
implicit procedure, and the method presented in this article can be
applied. The stationary Fokker-Planck equation considered for the test
is

\begin{equation} 
{\bf v} \cdot \nabla f + \dot{\bf v} \cdot \nabla_v f =  - \nabla_v f +  
\frac{1}{2} \sum_{i \neq j=1}^3 \frac{\partial^2 f}{\partial v_i \partial 
v_j} + \beta \sum_{i=1}^3 \frac{\partial^2 f}{\partial v_i^2}, 
\label{test} 
\end{equation}

\begin{figure} 
\epsfig{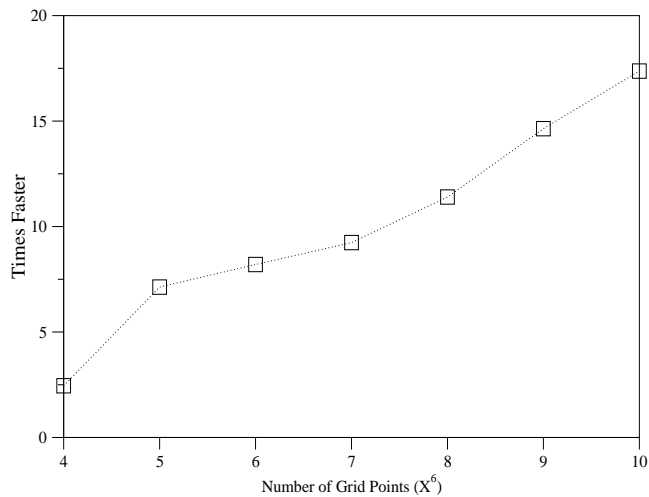}
\caption{Efficiency of our modified Stone method when compared  to the
{\it gmres} routine of Matlab that handles sparse matrices. When we 
increase the number of grid points our method becomes more efficient.
To handle the $9^6$ and $10^6$ grid in our computer, we lower the value
of the restarting parameter of the GMRES routine to the maximum possible
value that avoids computer memory problems.} \label{msmvsgmres}
\end{figure}

\noindent where $\dot{\bf v} =- \nabla \Phi$, $\Phi = 1/ \sqrt{x^2 +y^2
+z^2 +1^2}$, and $\beta$ is a constant. The form of Eq.  (\ref{test})
is similar to the equation found when Rosenbluth potentials
\cite{ros:mac} are used in a gravitational potential $\Phi$. Note that
the collision term has mixed velocity derivatives and that the
resulting coefficient matrix from the discretization is non symmetric.
We used central difference and a twenty-five point molecule to perform
the discretization of Eq. (\ref{test}), see also Table \ref{discret}.
Here, we find the solution of the above equation in a six dimensional
`box' of length 1.22 units. At the borders we used a Dirichlet boundary
condition of the form $f=\exp (-x^2 -y^2 -z^2) \exp (-v_x^2 -v_y^2
-v_z^2)$.

\begin{figure} 
\epsfig{width=10cm,height=7cm,file=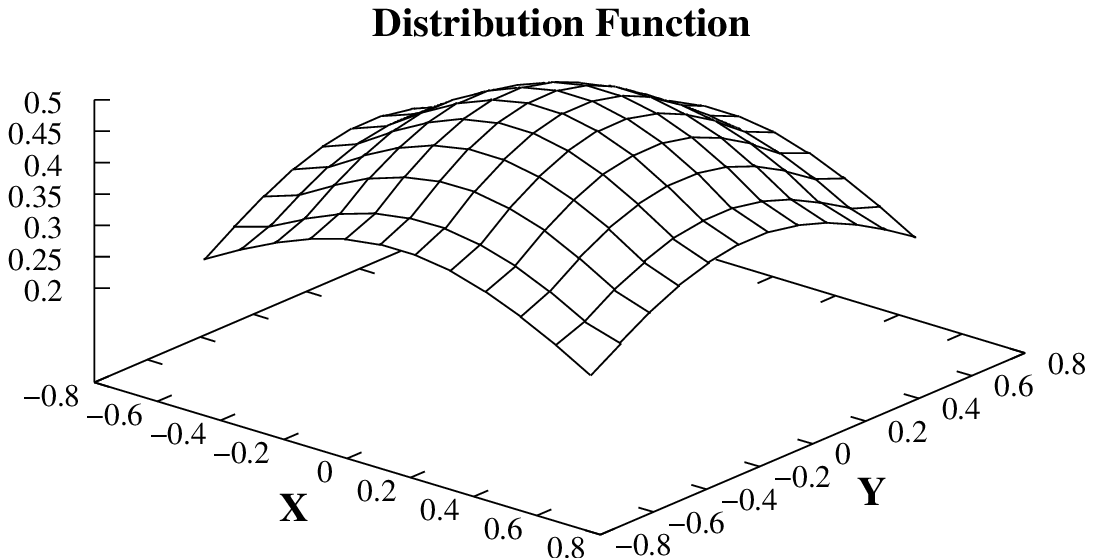}

\vspace{1cm} 
\epsfig{width=10cm,height=7cm,file=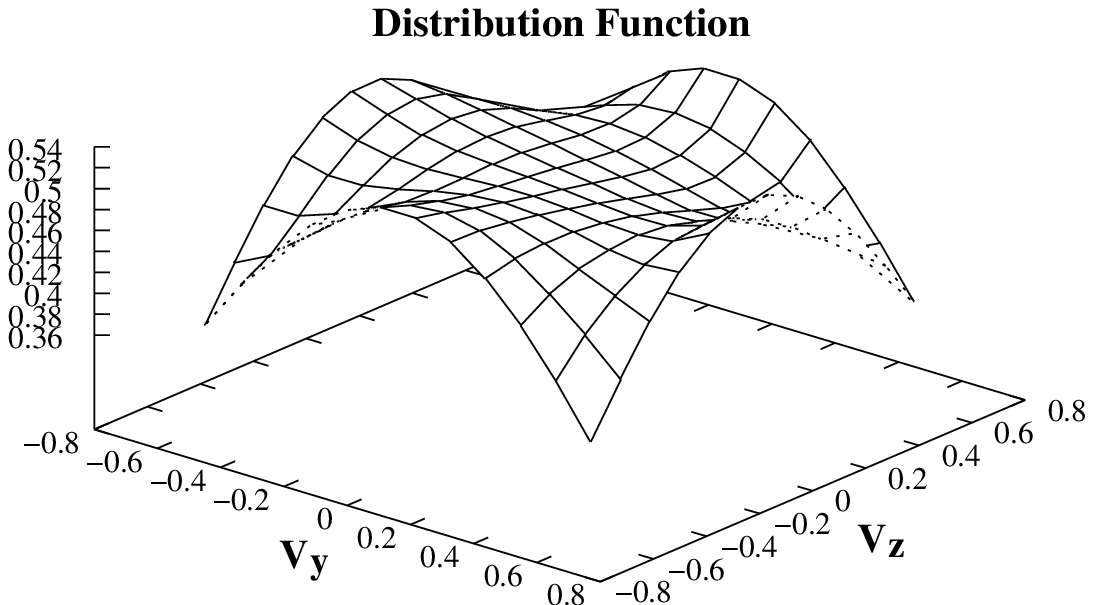}
\caption{Numerical solutions of the distribution function of Eq. 
(\ref{test}). Top: surface distribution for the case with $v_{x} 
\approx 0.38$, $v_y \approx -0.5$, $v_z \approx -0.5$ and $z \approx 
0.16$. Bottom: velocity distribution for the case with $x \approx -0.39$, 
$y \approx 0.06$, $z \approx -0.17$ and $v_x \approx 0.28$.} 
\label{distfig}
\end{figure}

We first started with coarse grid of four points per variable that
leads to a matrix of $4096^2$ elements (in this number we are not
considering the border grid points given by the boundary condition) and
$\beta=1$. We found for this case that our method spent approximately
0.2 seconds to find the solution (all the calculations were perform
with a Pentium IV of 1.8 GHz, the Fortran code was compiled with the
Linux free compiler). We stop the iteration when $\sum_{i=1}^n
(\Delta_{i}^{t+1}) < 10^{-10}$. The {\it gmres} routine and the public
GMRES code \cite{fra:gir} spent approximately 0.35 and 3 seconds
respectively to solve the system of equations with the same stop
criteria, but it could be more if we choose wrong the restarting
parameter. Here we are considering only the time spent to solve the
coefficient matrix and not the time due to create the coefficient
matrix and upload it into the code. In our code, we only upload 25
vectors of approximately $N_{node}$.

For a grid of five points per variable some code compilers can not
allow the storage of the coefficient matrix because is too large
$5^{12}$, approximately 244 millions of elements. This was the case of
the public code because it can not managed sparse matrices. For this
number of grid points the modified Stone method was almost 7 times
faster than the {\it gmres} routine. In Fig. \ref{msmvsgmres} we
present the efficiency of our method compare to the {\it gmres}
routine. Note that when we increase the number of grid points the
efficiency also increases. For a grid of nine points per variable the
{\it gmres} routine can not managed to find the solution in one
iteration because we have computer storage problems. To handle this, we
lower the restarting parameter to the maximum possible value that
avoids this problem. For a grid of ten point per variable the number of
elements in the coefficient matrix increases to $10^{12}$. Our code can
managed this huge amount of data in a faster and efficient way. The
time spent for this case was approximately 85 seconds. This should be
the time spent for each time step in a non stationary problem, which is
a great result considering the number of nodes in the grid and the
precision attained with the stop criteria. The variation of $\alpha$
between the accepted limits $0 < \alpha < 1$ slightly alter the time
spent in finding the solution. In Fig. \ref{distfig}, we present two
graphs of the distribution function from the numerical solution of Eq.
(\ref{test}) for a grid of ten grid per variable. One is the surface
distribution function on the plane $x-y$ and the other is the velocity
distribution function on the plane $v_y-v_z$; they are plotted at
different points of the grid. Note that these figures have not
Maxwellian form because Eq. (\ref{test}) is not in the collisional
regime for the parameters considered in the collision term. This
example was chosen only for didactic reasons.

Note that when we used central difference for discretization of Eq.  
(\ref{test}), the only contribution to the main diagonal comes from the
last right hand term, i.e. the term with $\beta$. For some values of
$\beta < 1$ our code diverge because our resulting coefficient matrix
is not diagonal dominant. For the same value of $\beta$, the {\it
gmres} routine converges. The break in convergence at these values of
$\beta$ coincide with the appearance of negatives values on the
distribution function solution found by {\it gmres}.  We know that one
of condition to attained a physical solution of the Fokker-Planck
equation is that the distribution function has to be always positive.
It is remarkable that for Eq. (\ref{test}) our code diverge when
physical solutions are not possible, this could be an indication that
we are applying a wrong scheme for discretization or that we are not
describing well the physical phenomena considered. As we said in the
Introduction, other discretization scheme may be applied to avoid this
problem. Thus, in our method the convergence is conditioned to the form
of the functions present in (\ref{fokker}) and (\ref{colterm}), i.e.
physical considerations; and to the difference scheme applied for
discretization, i.e. numerical implementation. For a non stationary
scheme the Crank-Nicolson decomposition in time is suggested because it
has implicit character and it is usually more stable than other
methods, but strictly speaking, the stability of the system has to be
studied for each particular case considered.

\section{Domains with curved boundaries}

\begin{figure} 
\epsfig{width=10cm,height=5cm,file=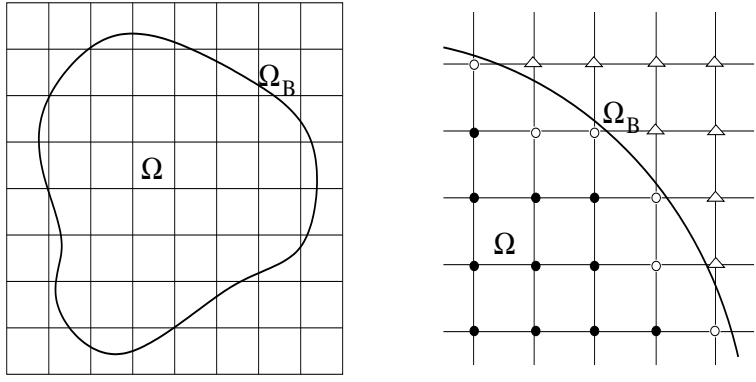}
\caption{Application of the incomplete $LU$ decomposition to a general
domain $\Omega$ with curved boundaries $\Omega_B$. The square structured
grid leads three classes of grid points: interior points (full circles),
boundary points (empty circles) and exterior points (triangles).}
\label{figgrid1}
\end{figure}

The $LU$ code was tested in the previous section with a
structure-orthogonal grid but this does not mean that it can not be
applied to more general geometries. To handle a non square domain we
proceed as follows. First we make a square structured grid with
$N_{node}=n_i n_j n_k n_l n_m n_n$ nodes, then we label them according
to the index $p$ defined in (\ref{p}). Later, the nodes that laid
outside the boundary regions are not considered for the calculations.
Now, lets see this procedure in an example. For didactic purpose we
used only the one dimensional case of the Fokker-Planck equation in
which we have two variables $(x,v_x)$. In Fig. \ref{figgrid1} is
depicted in the two dimensional plane $(x,v_x)$ a domain region
$\Omega$ with boundary $\Omega_B$. The domain $\Omega$ is filled with a
square structured grid that leads three classes of grid points: the
interior points in which the normal discretization procedure can be
done; the boundary points in which special care have to be taken when
Dirichlet, Neumann or mixed boundary conditions are applied; and the
exterior points that have to be neglected for the calculation, see Fig.
\ref{figgrid1}.

\begin{figure}
\epsfig{width=8cm,height=8cm,file=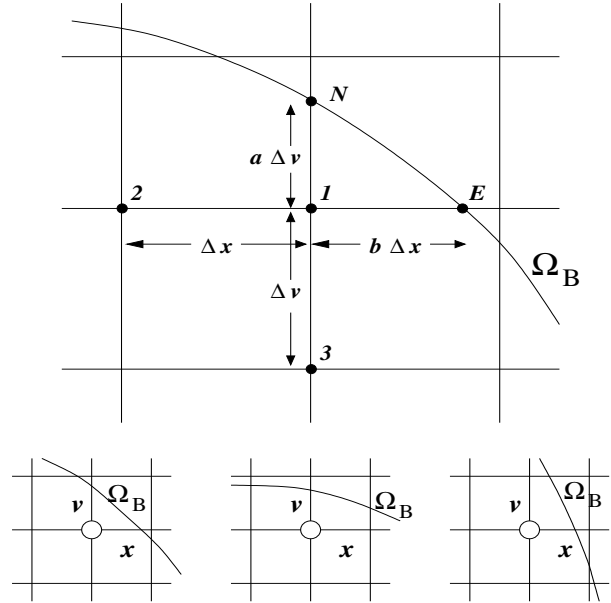}
\caption{Schematic representation of a five-point molecule for the finite
difference approximation of the derivatives in the plane $(x,v_x)$ for a
boundary point. Below, we see different boundary situations that can occur
in the numerical computation depending on the discretization grid and
boundary {\it S}.}
\label{figgrid2}
\end{figure}

The incorporation of the Dirichlet boundary conditions using central
differences for the boundary points of region $\Omega$ can be done
using a Taylor expansion around the nearest nodes. For example, in Fig.
\ref{figgrid2}, we take the nearest boundary points for point 1: the
internal node (node 2) and the point at the boundary $\Omega_B$ (point
$E$); and make two Taylor expansions around point 1. These expansions
give us (our conventions are: partial derivative with respect to the
coordinate $x$ denoted by ($,x$); $\Delta x$ is the discretization
interval in the $x$ direction),

\begin{eqnarray}
&&\Psi_E=\Psi_1+ b \Delta x \Psi_{,x} + \frac{(b \Delta x)^2}{2!} 
\Psi_{,xx} + O[\Delta x^3], \nonumber \\ 
&&\Psi_2=\Psi_1- \Delta x \Psi_{,x} + \frac{(\Delta x)^2}{2!}
\Psi_{,xx}+  O[\Delta x^3],
\end{eqnarray}

\noindent eliminating the second order derivative between these
equations we can express the partial derivative at a boundary point as,

\begin{equation}
\Psi_{,x}=\frac{1}{\Delta x} \left[ \frac{1}{b(1+b)} \Psi_E - 
\frac{b}{1+b} \Psi_2 - \frac{1-b}{b} \Psi_1 \right] + O[\Delta x^2]. 
\label{finitecentral}
\end{equation} 

\noindent When $b=1$ we recover the usual expression for the central
derivative. The same procedure can be applied to the northern boundary,
between node 3 and point $A$, and to the different types of boundary
nodes in Fig.  \ref{figgrid2}. Also, partial derivative of higher
orders and mixed derivatives can be found in a similar way.
Furthermore, it is also possible to implement Neumann boundary
conditions in an irregular boundary using finite difference, see for
instance \cite{for:was}.

The exterior points of Fig. \ref{figgrid1} are strictly necessary to
maintain the ordering of the nodes in a domain with curved boundaries.
This ordering is needed by the code to operate normally but they do not
enter into the calculations. To implement this condition, we have to
set the values of the elements $L^p_{[X]}$, $L^p_{P}$, $U^p_{[X]}$,
$R^p$, $\Upsilon^p$ and $\Delta^p$ of the external grid point (each
exterior point has a position $p$ from the one dimensional storage
index) equal to zero.

An application of the $LU$ method in a two dimensional Fokker-Planck
equation ($x,y,v_x,v_y$) with curved boundaries, as well as the
incomplete $LU$ decomposition for this case can be seen in
\cite{uje:let}.  Here, the distribution function of a stationary
gravitational thin disk is calculated. Note that in order to obtain the
elements of the $LU$ decomposition for the two dimensional case, we
have to start the calculations from the beginning, i.e. we can not used
the elements found in the three dimensional case.

\section{Concluding Remarks}

We have developed a variation of the incomplete $LU$ decomposition
proposed by Stone that solves the three dimensional linear
Fokker-Planck equation. The method presented can manage the large set
of linear equations that appears from the discretization procedure. The
convergence of the iterative process is done in a fast and effective
way. Also, this method can be easily adapted to support irregular
boundaries with Dirichlet, Neumann or mixed boundary conditions and can
be used to follow the evolution of a distribution function for non
stationary equations. In this case a Crank-Nicolson discretization in
time is recommended because it has implicit character and is more
stable than other methods, but strictly speaking, the stability of the
system has to be studied in each particular case considered. The good
properties that our method shares and the lack of methods that handle
the large amount of data (given by the Fokker-Planck equation with six
independent variables) make the method presented in this article worthy
and advantageous. In general, the algorithm presented in the article
can solve the system of equations that arises from the discretization
of the equation (\ref{fokker}) or similar equations (with or without
the mixed derivatives), but we have to keep in mind that its
convergence is conditioned to the form of the functions $\dot{\bf v}$,
$A({\bf x,v})$, $B({\bf x,v})$, $C({\bf x,v})$, $D({\bf x,v})$ and
$E_{ij}({\bf x,v})$ that appear in the collision term (\ref{colterm}).

\appendix

\section{Elements of $M=LU$ at node $p$} \label{MLU}

In this section we present how the elements of the matrix $M$ are
related to the elements of the upper $U$ and lower $L$ matrices.

\begin{eqnarray}
&&M^p_{DD3} = L^p_{DD3}, \nonumber \\
&&M^p_{DD3|N} = L^p_{DD3} U^{p-(n_{ijklm}+n_{ijkl})}_{N}, \nonumber \\
&&M^p_{DD3|E} = L^p_{DD3} U^{p-(n_{ijklm}+n_{ijkl})}_{E}, \nonumber \\
&&M^p_{DD3|T} = L^p_{DD3} U^{p-(n_{ijklm}+n_{ijkl})}_{T}, \nonumber \\
&&M^p_{DD3|U1} = L^p_{DD3} U^{p-(n_{ijklm}+n_{ijkl})}_{U1}, \nonumber \\
&&M^p_{DB3} = L^p_{DB3} + L^p_{DD3} U^{p-(n_{ijklm}+n_{ijkl})}_{UB2}, 
\nonumber \\
&&M^p_{DB3|N} = L^p_{DB3} U^{p-(n_{ijklm}+n_{ijk})}_{N}, \nonumber \\
&&M^p_{DB3|E} = L^p_{DB3} U^{p-(n_{ijklm}+n_{ijk})}_{E}, \nonumber \\
&&M^p_{DB3|T} = L^p_{DB3} U^{p-(n_{ijklm}+n_{ijk})}_{T}, \nonumber \\
&&M^p_{D3} = L^p_{D3} + L^p_{DB3} U^{p-(n_{ijklm}+n_{ijk})}_{U1} 
+ L^p_{DD3} U^{p-(n_{ijklm}+n_{ijkl})}_{U2}, \nonumber \\
&&M^p_{D3|N} = L^p_{D3} U^{p-n_{ijklm}}_{N}, \nonumber \\
&&M^p_{D3|E} = L^p_{D3} U^{p-n_{ijklm}}_{E}, \nonumber \\
&&M^p_{D3|T} = L^p_{D3} U^{p-n_{ijklm}}_{T}, \nonumber \\
&&M^p_{DT3} = L^p_{DT3} + L^p_{D3} U^{p-n_{ijklm}}_{U1} + L^p_{DD3} 
U^{p-(n_{ijklm}+n_{ijkl})}_{UT2}, \nonumber \\
&&M^p_{DT3|N} = L^p_{DT3} U^{p-(n_{ijklm}-n_{ijk})}_{N}, \nonumber \\
&&M^p_{DT3|E} = L^p_{DT3} U^{p-(n_{ijklm}-n_{ijk})}_{E}, \nonumber \\
&&M^p_{DT3|T} = L^p_{DT3} U^{p-(n_{ijklm}-n_{ijk})}_{T}, \nonumber \\
&&M^p_{DT3|U1} = L^p_{DT3} U^{p-(n_{ijklm}-n_{ijk})}_{U1}, \nonumber \\
&&M^p_{DB3|UB2} = L^p_{DB3} U^{p-(n_{ijklm}+n_{ijk})}_{UB2}, \nonumber \\
&&M^p_{D3|UB2} = L^p_{D3} U^{p-n_{ijklm}}_{UB2} + L^p_{DB3}
U^{p-(n_{ijklm}+n_{ijk})}_{U2}, \nonumber \\
&&M^p_{DU3} = L^p_{DU3} + L^p_{DT3} U^{p-(n_{ijklm}-n_{ijk})}_{UB2} +
L^p_{D3} U^{p-n_{ijklm}}_{U2} + L^p_{DB3} U^{p-(n_{ijklm}+n_{ijk})}_{UT2},
\nonumber \\
&&M^p_{DU3|N} = L^p_{DU3} U^{p-(n_{ijklm}-n_{ijkl})}_{N}, \nonumber \\
&&M^p_{DU3|E} = L^p_{DU3} U^{p-(n_{ijklm}-n_{ijkl})}_{E}, \nonumber \\
&&M^p_{DU3|T} = L^p_{DU3} U^{p-(n_{ijklm}-n_{ijkl})}_{T}, \nonumber \\
&&M^p_{DU3|U1} = L^p_{DU3} U^{p-(n_{ijklm}-n_{ijkl})}_{U1} + L^p_{DT3}
U^{p-(n_{ijklm}-n_{ijk})}_{U2} + L^p_{D3} U^{p-n_{ijklm}}_{UT2}, \nonumber
\\
&&M^p_{DT3|UT2} = L^p_{DT3} U^{p-(n_{ijklm}-n_{ijk})}_{UT2}, \nonumber \\
&&M^p_{DU3|UB2} = L^p_{DU3} U^{p-(n_{ijklm}-n_{ijkl})}_{UB2}, \nonumber \\
&&M^p_{DU3|U2} = L^p_{DU3} U^{p-(n_{ijklm}-n_{ijkl})}_{U2}, \nonumber \\
&&M^p_{DU3|UT2} = L^p_{DU3} U^{p-(n_{ijklm}-n_{ijkl})}_{UT2}, \nonumber \\
&&M^p_{DD3|UD3} = L^p_{DD3} U^{p-(n_{ijklm}+n_{ijkl})}_{UD3}, \nonumber \\
&&M^p_{DB2} = L^p_{DB2} + L^p_{DB3} U^{p-(n_{ijklm}+n_{ijk})}_{UD3} +
L^p_{DD3} U^{p-(n_{ijklm}+n_{ijkl})}_{UB3}, \nonumber \\
&&M^p_{DB2|N} = L^p_{DB2} U^{p-(n_{ijkl}+n_{ijk})}_{N}, \nonumber \\
&&M^p_{DB2|E} = L^p_{DB2} U^{p-(n_{ijkl}+n_{ijk})}_{E}, \nonumber \\
&&M^p_{DB2|T} = L^p_{DB2} U^{p-(n_{ijkl}+n_{ijk})}_{T}, \nonumber \\
&&M^p_{D2} = L^p_{D2} + L^p_{DB2} U^{p-(n_{ijkl}+n_{ijk})}_{U1} + L^p_{D3}
U^{p-n_{ijklm}}_{UD3} + L^p_{DD3} U^{p-(n_{ijklm}+n_{ijkl})}_{U3},
\nonumber \\
&&M^p_{D2|N} = L^p_{D2} U^{p-n_{ijkl}}_{N}, \nonumber \\
&&M^p_{D2|E} = L^p_{D2} U^{p-n_{ijkl}}_{E}, \nonumber \\
&&M^p_{D2|T} = L^p_{D2} U^{p-n_{ijkl}}_{T}, \nonumber \\
&&M^p_{DT2} = L^p_{DT2} + L^p_{D2} U^{p-n_{ijkl}}_{U1} + L^p_{DT3}
U^{p-(n_{ijklm}-n_{ijk})}_{UD3} + L^p_{DD3}
U^{p-(n_{ijklm}+n_{ijkl})}_{UT3}, \nonumber \\
&&M^p_{DT2|N} = L^p_{DT2} U^{p-(n_{ijkl}-n_{ijk})}_{N}, \nonumber \\
&&M^p_{DT2|E} = L^p_{DT2} U^{p-(n_{ijkl}-n_{ijk})}_{E}, \nonumber \\
&&M^p_{DT2|T} = L^p_{DT2} U^{p-(n_{ijkl}-n_{ijk})}_{T}, \nonumber \\
&&M^p_{DT2|U1} = L^p_{DT2} U^{p-(n_{ijkl}-n_{ijk})}_{U1}, \nonumber \\
&&M^p_{DB2|UB2} = L^p_{DB2} U^{p-(n_{ijkl}+n_{ijk})}_{UB2} + L^p_{DB3}
U^{p-(n_{ijklm}+n_{ijk})}_{UB3}, \nonumber \\
&&M^p_{D1} = L^p_{D1} + L^p_{D2} U^{p-n_{ijkl}}_{UB2} + L^p_{DB2}
U^{p-(n_{ijkl}+n_{ijk})}_{U2} + L^p_{D3} U^{p-n_{ijklm}}_{UB3} + L^p_{DB3}
U^{p-(n_{ijklm}+n_{ijk})}_{U3}, \nonumber \\
&&M^p_{D1|N} = L^p_{D1} U^{p-n_{ijk}}_{N}, \nonumber \\
&&M^p_{D1|E} = L^p_{D1} U^{p-n_{ijk}}_{E}, \nonumber \\
&&M^p_{D1|T} = L^p_{D1} U^{p-n_{ijk}}_{T}, \nonumber \\
&&M^p_{B} = L^p_{B}, \nonumber \\
&&M^p_{B|N} = L^p_{B} U^{p-n_{ij}}_{N}, \nonumber \\
&&M^p_{B|E} = L^p_{B} U^{p-n_{ij}}_{E}, \nonumber \\
&&M^p_{W} = L^p_{W}, \nonumber \\
&&M^p_{W|N} = L^p_{W} U^{p-n_{j}}_{N}, \nonumber \\
&&M^p_{S} = L^p_{S}, \nonumber \\
&&M^p_{P} = L^p_{P} + L^p_{S} U^{p-1}_{N} + L^p_{W} U^{p-n_j}_{E} + 
L^p_{B} U^{p-n_{ij}}_{T} + L^p_{D1} U^{p-n_{ijk}}_{U1} + L^p_{DT2} 
U^{p-(n_{ijkl}-n_{ijk})}_{UB2} + L^p_{D2} U^{p-n_{ijkl}}_{U2} \nonumber 
\\
&&\hspace{1cm}+ L^p_{DB2} U^{p-(n_{ijkl}+n_{ijk})}_{UT2} + L^p_{DU3} 
U^{p-(n_{ijklm}-n_{ijkl})}_{UD3} + L^p_{DT3} 
U^{p-(n_{ijklm}-n_{ijk})}_{UB3} + L^p_{D3} U^{p-n_{ijklm}}_{U3} \nonumber 
\\
&&\hspace{1cm}+ L^p_{DB3} U^{p-(n_{ijklm}+n_{ijk})}_{UT3} + L^p_{DD3} 
U^{p-(n_{ijklm}+n_{ijkl})}_{UU3}, \nonumber \\
&&M^p_{N} = L^p_{P} U^p_{N}, \nonumber \\
&&M^p_{S|E} = L^p_{S} U^{p-1}_{E}, \nonumber \\
&&M^p_{E} = L^p_{P} U^p_{E}, \nonumber \\
&&M^p_{W|T} = L^p_{W} U^{p-n_j}_{T}, \nonumber \\
&&M^p_{S|T} = L^p_{S} U^{p-1}_{T}, \nonumber \\
&&M^p_{T} = L^p_{P} U^p_{T}, \nonumber \\
&&M^p_{B|U1} = L^p_{B} U^{p-n_{ij}}_{U1}, \nonumber \\
&&M^p_{W|U1} = L^p_{W} U^{p-n_j}_{U1}, \nonumber \\
&&M^p_{S|U1} = L^p_{S} U^{p-1}_{U1}, \nonumber \\
&&M^p_{U1} = L^p_{P} U^p_{U1} + L^p_{DT2} U^{p-(n_{ijkl}-n_{ijk})}_{U2} + 
L^p_{D2} U^{p-n_{ijkl}}_{UT2} + L^p_{DT3} U^{p-(n_{ijklm}-n_{ijk})}_{U3} + 
L^p_{D3} U^{p-n_{ijklm}}_{UT3}, \nonumber \\
&&M^p_{DT2|UT2} =L^p_{DT2} U^{p-(n_{ijkl}-n_{ijk})}_{UT2} + L^p_{DT3}
U^{p-(n_{ijklm}-n_{ijk})}_{UT3}, \nonumber \\
&&M^p_{D1|UB2} = L^p_{D1} U^{p-n_{ijk}}_{UB2}, \nonumber \\
&&M^p_{B|UB2} = L^p_{B} U^{p-n_{ij}}_{UB2}, \nonumber \\
&&M^p_{W|UB2} = L^p_{W} U^{p-n_j}_{UB2}, \nonumber \\
&&M^p_{S|UB2} = L^p_{S} U^{p-1}_{UB2}, \nonumber \\
&&M^p_{UB2} = L^p_{P} U^p_{UB2} + L^p_{D1} U^{p-n_{ijk}}_{U2} + L^p_{DU3}
U^{p-(n_{ijklm}-n_{ijkl})}_{UB3} + L^p_{DB3}
U^{p-(n_{ijklm}+n_{ijk})}_{UU3}, \nonumber \\
&&M^p_{B|U2} = L^p_{B} U^{p-n_{ij}}_{U2}, \nonumber \\
&&M^p_{W|U2} = L^p_{W} U^{p-n_j}_{U2}, \nonumber \\
&&M^p_{S|U2} = L^p_{S} U^{p-1}_{U2}, \nonumber \\
&&M^p_{U2} = L^p_{P} U^p_{U2} + L^p_{D1} U^{p-n_{ijk}}_{UT2} + L^p_{DU3}
U^{p-(n_{ijklm}-n_{ijkl})}_{U3} + L^p_{D3} U^{p-n_{ijklm}}_{UU3},
\nonumber \\
&&M^p_{B|UT2} = L^p_{B} U^{p-n_{ij}}_{UT2}, \nonumber \\
&&M^p_{W|UT2} = L^p_{W} U^{p-n_j}_{UT2}, \nonumber \\
&&M^p_{S|UT2} = L^p_{S} U^{p-1}_{UT2}, \nonumber \\
&&M^p_{UT2} = L^p_{P} U^p_{UT2} + L^p_{DU3} 
U^{p-(n_{ijklm}-n_{ijkl})}_{UT3} + L^p_{DT3} 
U^{p-(n_{ijklm}-n_{ijk})}_{UU3}, \nonumber \\
&&M^p_{DU3|UU3} = L^p_{DU3} U^{p-(n_{ijklm}-n_{ijkl})}_{UU3}, \nonumber \\
&&M^p_{DB2|UD3} = L^p_{DB2} U^{p-(n_{ijkl}+n_{ijk})}_{UD3}, \nonumber \\
&&M^p_{D2|UD3} = L^p_{D2} U^{p-n_{ijkl}}_{UD3}, \nonumber \\
&&M^p_{DT2|UD3} = L^p_{DT2} U^{p-(n_{ijkl}-n_{ijk})}_{UD3}, \nonumber \\
&&M^p_{DB2|UB3} = L^p_{DB2} U^{p-(n_{ijkl}+n_{ijk})}_{UB3}, \nonumber \\
&&M^p_{D1|UD3} = L^p_{D1} U^{p-n_{ijk}}_{UD3} + L^p_{D2}
U^{p-n_{ijkl}}_{UB3} + L^p_{DB2} U^{p-(n_{ijkl}+n_{ijk})}_{U3}, \nonumber
\\
&&M^p_{B|UD3} = L^p_{B} U^{p-n_{ij}}_{UD3}, \nonumber \\
&&M^p_{W|UD3} = L^p_{W} U^{p-n_j}_{UD3}, \nonumber \\
&&M^p_{S|UD3} = L^p_{S} U^{p-1}_{UD3}, \nonumber \\
&&M^p_{UD3} = L^p_{P} U^p_{UD3} + L^p_{DT2} U^{p-(n_{ijkl}-n_{ijk})}_{UB3}
+ L^p_{D2} U^{p-n_{ijkl}}_{U3} + L^p_{DB2} U^{p-(n_{ijkl}+n_{ijk})}_{UT3},
\nonumber \\
&&M^p_{DT2|U3} = L^p_{DT2} U^{p-(n_{ijkl}-n_{ijk})}_{U3} + L^p_{D2}
U^{p-n_{ijkl}}_{UT3}, \nonumber \\
&&M^p_{DT2|UT3} = L^p_{DT2} U^{p-(n_{ijkl}-n_{ijk})}_{UT3}, \nonumber \\
&&M^p_{D1|UB3} = L^p_{D1} U^{p-n_{ijk}}_{UB3}, \nonumber \\
&&M^p_{B|UB3} = L^p_{B} U^{p-n_{ij}}_{UB3}, \nonumber \\
&&M^p_{W|UB3} = L^p_{W} U^{p-n_j}_{UB3}, \nonumber \\
&&M^p_{S|UB3} = L^p_{S} U^{p-1}_{UB3}, \nonumber \\
&&M^p_{UB3} = L^p_{P} U^p_{UB3} + L^p_{D1} U^{p-n_{ijk}}_{U3} + L^p_{DB2} 
U^{p-(n_{ijkl}+n_{ijk})}_{UU3}, \nonumber \\
&&M^p_{B|U3} = L^p_{B} U^{p-n_{ij}}_{U3}, \nonumber \\
&&M^p_{W|U3} = L^p_{W} U^{p-n_j}_{U3}, \nonumber \\
&&M^p_{S|U3} = L^p_{S} U^{p-1}_{U3}, \nonumber \\
&&M^p_{U3} = L^p_{P} U^p_{U3} + L^p_{D1} U^{p-n_{ijk}}_{UT3} + L^p_{D2} 
U^{p-n_{ijkl}}_{UU3}, \nonumber \\
&&M^p_{B|UT3} = L^p_{B} U^{p-n_{ij}}_{UT3}, \nonumber \\
&&M^p_{W|UT3} = L^p_{W} U^{p-n_j}_{UT3}, \nonumber \\
&&M^p_{S|UT3} = L^p_{S} U^{p-1}_{UT3}, \nonumber \\
&&M^p_{UT3} = L^p_{P} U^p_{UT3} + L^p_{DT2} 
U^{p-(n_{ijkl}-n_{ijk})}_{UU3}, \nonumber \\
&&M^p_{D1|UU3} = L^p_{D1} U^{p-n_{ijk}}_{UU3}, \nonumber \\
&&M^p_{B|UU3} = L^p_{B} U^{p-n_{ij}}_{UU3}, \nonumber \\
&&M^p_{W|UU3} = L^p_{W} U^{p-n_j}_{UU3}, \nonumber \\
&&M^p_{S|UU3} = L^p_{S} U^{p-1}_{UU3}, \nonumber \\
&&M^p_{UU3} = L^p_{P} U^p_{UU3}.
\label{relaM} 
\end{eqnarray}

\section{Explicit Form of Eq. (\ref{Oapprox})} \label{Oexplicit}

In this section we present the explicit form of $O \Psi \approx 0$ for
each grid node.

\begin{eqnarray}
&&O_{DD3} \Psi_{DD3} + O_{DB3} \Psi_{DB3} + O_{D3} \Psi_{D3} + O_{DT3}
\Psi_{DT3} + O_{DU3} \Psi_{DU3} + O_{DB2} \Psi_{DB2} + O_{D2} \Psi_{D2} +
O_{DT2} \Psi_{DT2} \nonumber \\
&&+ O_{D1} \Psi_{D1} +O_{B} \Psi_{B} +O_{W} 
\Psi_{W} +O_{S} \Psi_{S} +O_{P} \Psi_{P} +O_{N} \Psi_{N} +O_{E} \Psi_{E} 
+O_{T} \Psi_{T} + O_{U1} \Psi_{U1} + O_{UB2} \Psi_{UB2} \nonumber \\
&&+ O_{U2} \Psi_{U2} + O_{UT2} \Psi_{UT2} + O_{UD3} \Psi_{UD3} + O_{UB3} 
\Psi_{UB3} + O_{U3} \Psi_{U3} + O_{UT3} \Psi_{UT3} + O_{UU3} \Psi_{UU3} + 
M_{DD3|N} \Psi_{DD3|N} \nonumber \\
&&+ M_{DD3|E} \Psi_{DD3|E} + M_{DD3|T} \Psi_{DD3|T} + M_{DD3|U1} 
\Psi_{DD3|U1} + M_{DB3|N} \Psi_{DB3|N} + M_{DB3|E} \Psi_{DB3|E} \nonumber 
\\
&&+ M_{DB3|T} \Psi_{DB3|T} + M_{D3|N} \Psi_{D3|N} + M_{D3|E} \Psi_{D3|E} + 
M_{D3|T} \Psi_{D3|T} + M_{DT3|N} \Psi_{DT3|N} + M_{DT3|E} \Psi_{DT3|E} 
\nonumber \\
&&+ M_{DT3|T} \Psi_{DT3|T} + M_{DT3|U1} \Psi_{DT3|U1} + M_{DB3|UB2} 
\Psi_{DB3|UB2} + M_{DB3|U2} \Psi_{DB3|U2} + M_{D3|UB2} \Psi_{D3|UB2} 
\nonumber \\
&&+ M_{DU3|N} \Psi_{DU3|N} + M_{DU3|E} \Psi_{DU3|E} + M_{DU3|T} 
\Psi_{DU3|T} + M_{D3|UT2} \Psi_{D3|UT2} + M_{DT3|U2} \Psi_{DT3|U2} 
\nonumber \\
&&+ M_{DU3|U1} \Psi_{DU3|U1} + M_{DT3|UT2} \Psi_{DT3|UT2} + M_{DU3|UB2} 
\Psi_{DU3|UB2} + M_{DU3|U2} \Psi_{DU3|U2} + M_{DU3|UT2} \Psi_{DU3|UT2} 
\nonumber \\
&&+ M_{DD3|UD3} \Psi_{DD3|UD3} + M_{DB2|N} \Psi_{DB2|N} + M_{DB2|E} 
\Psi_{DB2|E} + M_{DB2|T} \Psi_{DB2|T} + M_{D2|N} \Psi_{D2|N} + M_{D2|E} 
\Psi_{D2|E} \nonumber \\
&&+ M_{D2|T} \Psi_{D2|T} + M_{DT2|N} \Psi_{DT2|N} + M_{DT2|E} \Psi_{DT2|E} 
+ M_{DT2|T} \Psi_{DT2|T} + M_{DT2|U1} \Psi_{DT2|U1} \nonumber \\
&&+ M_{DB3|UB3} \Psi_{DB3|UB3} + M_{DB2|UB2} \Psi_{DB2|UB2} + M_{D1|N} 
\Psi_{D1|N} + M_{D1|E} \Psi_{D1|E} + M_{D1|T} \Psi_{D1|T} + M_{B|N} 
\Psi_{B|N} \nonumber \\
&&+ M_{B|E} \Psi_{B|E} + M_{W|N} \Psi_{W|N} + M_{S|E} \Psi_{S|E} + 
M_{W|T} \Psi_{W|T} + M_{S|T} \Psi_{S|T} + M_{B|U1} \Psi_{B|U1} + M_{W|U1} 
\Psi_{W|U1} \nonumber \\
&&+ M_{S|U1} \Psi_{S|U1} + M_{DT2|UT2} \Psi_{DT2|UT2} + M_{DT3|UT3} 
\Psi_{DT3|UT3} +M_{D1|UB2} \Psi_{D1|UB2} + M_{B|UB2} \Psi_{B|UB2} 
\nonumber \\
&&+ M_{W|UB2} \Psi_{W|UB2} + M_{S|UB2} \Psi_{S|UB2} + M_{B|U2} \Psi_{B|U2} + 
M_{W|U2} \Psi_{W|U2} + M_{S|U2} \Psi_{S|U2} + M_{B|UT2} \Psi_{B|UT2} 
\nonumber \\
&&+ M_{W|UT2} \Psi_{W|UT2} + M_{S|UT2} \Psi_{S|UT2} + M_{DU3|UU3} 
\Psi_{DU3|UU3} + M_{DB2|UD3} \Psi_{DB2|UD3} + M_{D2|UD3} \Psi_{D2|UD3} 
\nonumber \\
&&+ M_{DT2|UD3} \Psi_{DT2|UD3} + M_{DB2|UB3} \Psi_{DB2|UB3} + M_{D1|UD3} 
\Psi_{D1|UD3} + M_{D2|UB3} \Psi_{D2|UB3} + M_{DB2|U3} \Psi_{DB2|U3} 
\nonumber \\
&&+ M_{B|UD3} \Psi_{B|UD3} + M_{W|UD3} \Psi_{W|UD3} + 
M_{S|UD3} \Psi_{S|UD3} + M_{DT2|U3} \Psi_{DT2|U3} + M_{D2|UT3} 
\Psi_{D2|UT3} \nonumber \\
&&+ M_{DT2|UT3} \Psi_{DT2|UT3} + M_{D1|UB3} \Psi_{D1|UB3} + M_{B|UB3} 
\Psi_{B|UB3} + M_{W|UB3} \Psi_{W|UB3} + M_{S|UB3} \Psi_{S|UB3} \nonumber 
\\
&&+ M_{B|U3} \Psi_{B|U3} + M_{W|U3} \Psi_{W|U3} + M_{S|U3} \Psi_{S|U3} + 
M_{B|UT3} \Psi_{B|UT3} + M_{W|UT3} \Psi_{W|UT3} + M_{S|UT3} \Psi_{S|UT3} 
\nonumber \\
&&+ M_{D1|UU3} \Psi_{D1|UU3} + M_{B|UU3} \Psi_{B|UU3} + M_{W|UU3} 
\Psi_{W|UU3} + M_{S|UU3} \Psi_{S|UU3} \approx 0.
\label{noderelation}
\end{eqnarray}

\section{$O$ as a Linear Combination of $M$} \label{linear}

In this section we present the elements of the matrix $O$ as a linear
combination of the elements of matrix $M$ using the relations
(\ref{combination}).

\begin{eqnarray}
&&O_{DD3} = - \alpha (M_{DD3|N} + M_{DD3|E} + M_{DD3|T} + M_{DD3|U1} + 
M_{DD3|UD3}), \nonumber \\
&&O_{DB3} = - \alpha (M_{DB3|N} + M_{DB3|E} + M_{DB3|T} + M_{DB3|UB2} + 
M_{DB3|U2} + M_{DB3|UB3}), \nonumber \\
&&O_{D3} = - \alpha (M_{D3|N} + M_{D3|E} + M_{D3|T} + M_{D3|UB2} + 
M_{D3|UT2}), \nonumber \\
&&O_{DT3} = -\alpha (M_{DT3|N} + M_{DT3|E} + M_{DT3|T} + M_{DT3|U1} + 
M_{DT3|U2} + M_{DT3|UT2} + M_{DT3|UT3}), \nonumber \\
&&O_{DU3} = -\alpha (M_{DU3|N} + M_{DU3|E} + M_{DU3|T} + M_{DU3|U1} +
M_{DU3|UB2} + M_{DU3|U2} + M_{DU3|UT2} + M_{DU3|UU3}), \nonumber \\
&&O_{DB2} = - \alpha (M_{DB2|N} + M_{DB2|E} + M_{DB2|T} + M_{DB2|UB2} +
M_{DB2|UD3} + M_{DB2|UB3} + M_{DB2|U3}), \nonumber \\
&&O_{D2} = - \alpha (M_{D2|N} + M_{D2|E} + M_{D2|T} + M_{D2|UD3} + 
M_{D2|UB3} + M_{D2|UT3}), \nonumber \\
&&O_{DT2} = - \alpha (M_{DT2|N} + M_{DT2|E} + M_{DT2|T} + M_{DT2|U1} +
M_{DT2|UT2} + M_{DT2|UD3} + M_{DT2|U3} + M_{DT2|UT3}), \nonumber \\
&&O_{D1} = - \alpha (M_{D1|N} + M_{D1|E} + M_{D1|T} + M_{D1|UB2} +
M_{D1|UD3} + M_{D1|UB3} + M_{D1|UU3}), \nonumber \\
&&O_{B} = -\alpha (M_{B|N} + M_{B|E} + M_{B|U1} + M_{B|UB2} + M_{B|U2} + 
M_{B|UT2} + M_{B|UD3} + M_{B|UB3} + M_{B|U3} + M_{B|UT3} \nonumber \\
&&\hspace{1cm}+ M_{B|UU3}), \nonumber \\
&&O_{W} = -\alpha (M_{W|N} + M_{W|T} + M_{W|U1} + M_{W|UB2} + M_{W|U2} + 
M_{W|UT2} + M_{W|UD3} + M_{W|UB3} + M_{W|U3} + M_{W|UT3} \nonumber \\
&&\hspace{1cm}+ M_{W|UU3}), \nonumber \\
&&O_{S} = -\alpha (M_{S|E} + M_{S|T} + M_{S|U1} + M_{S|UB2} + M_{S|U2} + 
M_{S|UT2} + M_{S|UD3} + M_{S|UB3} + M_{S|U3} + M_{S|UT3} \nonumber \\
&&\hspace{1cm}+ M_{S|UU3}), \nonumber \\
&&O_{P} = \alpha (M_{DD3|N} + M_{DD3|E} + M_{DD3|T} + M_{DD3|U1} +
M_{DB3|N} + M_{DB3|E} + M_{DB3|T} + M_{D3|N} + M_{D3|E} + M_{D3|T} 
\nonumber \\
&&\hspace{1cm}+ M_{DT3|N} + M_{DT3|E} + M_{DT3|T} + M_{DT3|U1} + 
M_{DB3|UB2} + M_{DB3|U2} + M_{D3|UB2} + M_{DU3|N} + M_{DU3|E} \nonumber \\
&&\hspace{1cm}+ M_{DU3|T} + M_{D3|UT2} + M_{DT3|U2} + M_{DU3|U1} + 
M_{DT3|UT2} + M_{DU3|UB2} + M_{DU3|U2} + M_{DU3|UT2} \nonumber \\
&&\hspace{1cm}+ M_{DD3|UD3} + M_{DB2|N} + M_{DB2|E} + M_{DB2|T} + 
M_{D2|N} + M_{D2|E}+ M_{D2|T} + M_{DT2|N} + M_{DT2|E} + M_{DT2|T} 
\nonumber \\
&&\hspace{1cm}+ M_{DT2|U1} + M_{DB3|UB3} + M_{DB2|UB2} + M_{D1|N} + 
M_{D1|E} + M_{D1|T} + M_{B|N} + M_{B|E} + M_{W|N} + M_{S|E} \nonumber \\
&&\hspace{1cm}+ M_{W|T} + M_{S|T} + M_{B|U1} + M_{W|U1} + M_{S|U1} + 
M_{DT2|UT2} + M_{DT3|UT3} + M_{D1|UB2} + M_{B|UB2} + M_{W|UB2} \nonumber 
\\
&&\hspace{1cm}+ M_{S|UB2} + M_{B|U2} + M_{W|U2} + M_{S|U2} + M_{B|UT2} + 
M_{W|UT2} + M_{S|UT2} + M_{DU3|UU3} + M_{DB2|UD3} \nonumber \\
&&\hspace{1cm}+ M_{D2|UD3} + M_{DT2|UD3} + M_{DB2|UB3} + M_{D1|UD3} + 
M_{D2|UB3} + M_{DB2|U3} + M_{B|UD3} + M_{W|UD3} + M_{S|UD3} \nonumber \\
&&\hspace{1cm}+ M_{DT2|U3} + M_{D2|UT3} + M_{DT2|UT3} + M_{D1|UB3} + 
M_{B|UB3} + M_{W|UB3} + M_{S|UB3} + M_{B|U3} + M_{W|U3} \nonumber \\
&&\hspace{1cm}+ M_{S|U3} + M_{B|UT3} + M_{W|UT3} + M_{S|UT3} + M_{D1|UU3} 
+ M_{B|UU3} + M_{W|UU3} + M_{S|UU3}), \nonumber \\
&&O_{N} = - \alpha (M_{DD3|N} + M_{DB3|N} + M_{D3|N} + M_{DT3|N} + 
M_{DU3|N} + M_{DB2|N} + M_{D2|N} + M_{DT2|N} + M_{D1|N} + M_{B|N} 
\nonumber \\
&&\hspace{1cm}+ M_{W|N}), \nonumber \\
&&O_{E} = - \alpha (M_{DD3|E} + M_{DB3|E} + M_{D3|E} + M_{DT3|E} + 
M_{DU3|E} + M_{DB2|E} + M_{D2|E} + M_{DT2|E} + M_{D1|E} + M_{B|E} 
\nonumber \\
&&\hspace{1cm}+ M_{S|E}), \nonumber \\
&&O_{T} = -\alpha (M_{DD3|T} + M_{DB3|T} + M_{D3|T} + M_{DT3|T} + 
M_{DU3|T} + M_{DB2|T} + M_{D2|T} + M_{DT2|T} + M_{D1|T} + M_{W|T} 
\nonumber \\
&&\hspace{1cm}+ M_{S|T}), \nonumber \\
&&O_{U1} = - \alpha (M_{DD3|U1} + M_{DT3|U1} + M_{DU3|U1} + M_{DT2|U1} + 
M_{B|U1} + M_{W|U1} + M_{S|U1}), \nonumber \\
&&O_{UB2} = - \alpha (M_{DB3|UB2} + M_{D3|UB2} + M_{DU3|UB2} + 
M_{DB2|UB2} + M_{D1|UB2} + M_{B|UB2} + M_{W|UB2} + M_{S|UB2}), \nonumber 
\\
&&O_{U2} = - \alpha (M_{DB3|U2} + M_{DT3|U2} + M_{DU3|U2} + M_{B|U2} + 
M_{W|U2} + M_{S|U2}), \nonumber \\ 
&&O_{UT2} = - \alpha (M_{D3|UT2} + M_{DT3|UT2} + M_{DU3|UT2} + 
M_{DT2|UT2} + M_{B|UT2} + M_{W|UT2} + M_{S|UT2}), \nonumber \\
&&O_{UD3} = - \alpha (M_{DD3|UD3} + M_{DB2|UD3} + M_{D2|UD3} + M_{DT2|UD3} 
+ M_{D1|UD3} + M_{B|UD3} + M_{W|UD3} + M_{S|UD3}), \nonumber \\
&&O_{UB3} = - \alpha (M_{DB3|UB3} + M_{DB2|UB3} + M_{D2|UB3} + M_{D1|UB3} 
+ M_{B|UB3} + M_{W|UB3} + M_{S|UB3}), \nonumber \\
&&O_{U3} = -\alpha (M_{DB2|U3} + M_{DT2|U3} + M_{B|U3} + M_{W|U3} + 
M_{S|U3}), \nonumber \\
&&O_{UT3} = - \alpha (M_{DT3|UT3} + M_{D2|UT3} + M_{DT2|UT3} + M_{B|UT3} + 
M_{W|UT3} + M_{S|UT3}), \nonumber \\
&&O_{UU3} = - \alpha (M_{DU3|UU3} + M_{D1|UU3} + M_{B|UU3} + M_{W|UU3} + 
M_{S|UU3}). 
\label{relaO}
\end{eqnarray}

\section{Explicit form of the functions $C_{[X]}$ and $K_{[X]}$}
\label{CandK}

In this section we show the functions $C_{[X]}$ and $K_{[X]}$ that
appear in the final expression (\ref{LUdecom}) of the $LU$
decomposition.

\begin{eqnarray}
&&C_{DB3} = L_{DD3}^p U_{UB2}^{p - (n_{ijklm} + n_{ijkl})}, \nonumber \\
&&C_{D3} = L_{DD3}^p U_{U2}^{p - (n_{ijklm} + n_{ijkl})} + L_{DB3}^p 
U_{U1}^{p - (n_{ijklm} + n_{ijk})}, \nonumber \\
&&C_{DT3} = L_{DD3}^p U_{UT2}^{p - (n_{ijklm} + n_{ijkl})} + L_{D3}^p 
U_{U1}^{p - n_{ijklm}}, \nonumber \\
&&C_{DU3} = L_{DB3}^p U_{UT2}^{p - (n_{ijklm} + n_{ijk})} + L_{D3}^p
U_{U2}^{p - n_{ijklm}} + L_{DT3}^p U_{UB2}^{p - (n_{ijklm} - n_{ijk})}, 
\nonumber \\
&&C_{DB2} = L_{DD3}^p U_{UB3}^{p - (n_{ijklm} + n_{ijkl})} + L_{DB3}^p 
U_{UD3}^{p - (n_{ijklm} + n_{ijk})}, \nonumber \\
&&C_{D2} = L_{DD3}^p U_{U3}^{p - (n_{ijklm} + n_{ijkl})} + L_{D3}^p
U_{UD3}^{p - n_{ijklm}} + L_{DB2}^p U_{U1}^{p - (n_{ijkl} + n_{ijk})}, 
\nonumber \\
&&C_{DT2} = L_{DD3}^p U_{UT3}^{p - (n_{ijklm} + n_{ijkl})} + L_{DT3}^p
U_{UD3}^{p - (n_{ijklm} - n_{ijk})} + L_{D2}^p U_{U1}^{p - n_{ijkl}}, 
\nonumber \\
&&C_{D1} = L_{DB3}^p U_{U3}^{p - (n_{ijklm} + n_{ijk})} + L_{D3}^p
U_{UB3}^{p - n_{ijklm}} + L_{DB2}^p U_{U2}^{p - (n_{ijkl} + n_{ijk})} +
L_{D2}^p U_{UB2}^{p - n_{ijkl}}, \nonumber \\
&&C_{U1} = L_{DT2}^p U_{U2}^{p- (n_{ijkl} - n_{ijk})} + L_{D2}^p
U_{UT2}^{p- n_{ijkl}} + L_{DT3}^p U_{U3}^{p - (n_{ijklm} - n_{ijk})} +
L_{D3}^p U_{UT3}^{p -n_{ijklm}}, \nonumber \\
&&C_{UB2} = L_{D1}^p U_{U2}^{p - n_{ijk}} + L_{DU3}^p U_{UB3}^{p -
(n_{ijklm} - n_{ijkl})} + L_{DB3}^p U_{UU3}^{p -( n_{ijklm} + n_{ijk})},
\nonumber \\
&&C_{U2} = L_{D1}^p U_{UT2}^{p - n_{ijk}} + L^p_{DU3} U_{U3}^{p -
(n_{ijklm} - n_{ijkl})} + L_{D3}^p U_{UU3}^{p - n_{ijklm}}, \nonumber \\
&&C_{UT2} = L_{DU3}^p U_{UT3}^{p - (n_{ijklm} - n_{ijkl})} + L_{DT3}^p
U_{UU3}^{p- (n_{ijklm} - n_{ijk})}, \nonumber \\
&&C_{UD3} = L_{DT2}^p U_{UB3}^{p - (n_{ijkl} - n_{ijk})} + L_{D2}^p
U_{U3}^{p - n_{ijkl}} + L_{DB2}^p U_{UT3}^{p - (n_{ijkl} + n_{ijk})},
\nonumber \\
&&C_{UB3} = L_{D1}^p U_{U3}^{p - n_{ijk}} + L_{DB2}^p U_{UU3}^{p -
(n_{ijkl} + n_{ijk})}, \nonumber \\
&&C_{U3} = L_{D1}^p U_{UT3}^{p -n_{ijk}} + L_{D2}^p U_{UU3}^{p -
n_{ijkl}}, \nonumber \\
&&C_{UT3} = L_{DT2}^p U_{UU3}^{p - (n_{ijkl} - n_{ijk})}, \nonumber \\
&&C_{DD3} = C_{B} = C_{W} = C_{S} = C_{N} = C_{E} = C_{T} = C_{UU3} = 0, 
\nonumber \\
&&K_{DD3} = U_{N}^{p-(n_{ijklm} + n_{ijkl})} + U_{E}^{p-(n_{ijklm} + 
n_{ijkl})} + U_{T}^{p-(n_{ijklm} + n_{ijkl})} + U_{U1}^{p-(n_{ijklm} +
n_{ijkl})} + U_{UD3}^{p-(n_{ijklm} + n_{ijkl})}, \nonumber \\
&&K_{DB3} = U_{N}^{p-(n_{ijklm} + n_{ijk})} + U_{E}^{p-(n_{ijklm} + 
n_{ijk})} + U_{T}^{p-(n_{ijklm} + n_{ijk})} + U_{UB2}^{p-(n_{ijklm} + 
n_{ijk})} + U_{U2}^{p-(n_{ijklm} + n_{ijk})} \nonumber \\
&&\hspace{1cm}+ U_{UB3}^{p-(n_{ijklm} + n_{ijk})}, \nonumber \\
&&K_{D3} = U_{N}^{p - n_{ijklm}} + U_{E}^{p - n_{ijklm}} + U_{T}^{p -
n_{ijklm}} + U_{UB2}^{p - n_{ijklm}} + U_{UT2}^{p - n_{ijklm}}, \nonumber 
\\
&&K_{DT3} = U_{N}^{p- (n_{ijklm} - n_{ijk})} + U_{E}^{p- (n_{ijklm} -
n_{ijk})} + U_{T}^{p- (n_{ijklm} - n_{ijk})} + U_{U1}^{p- (n_{ijklm} -
n_{ijk})} + U_{U2}^{p- (n_{ijklm} - n_{ijk})} \nonumber \\
&&\hspace{1cm}+ U_{UT2}^{p- (n_{ijklm} - n_{ijk})} + U_{UT3}^{p- 
(n_{ijklm} - n_{ijk})}, \nonumber \\
&&K_{DU3} = U_{N}^{p-( n_{ijklm} - n_{ijkl})} + U_{E}^{p-( n_{ijklm} -
n_{ijkl})} + U_{T}^{p-( n_{ijklm} - n_{ijkl})} + U_{U1}^{p-( n_{ijklm} -
n_{ijkl})} + U_{UB2}^{p-( n_{ijklm} - n_{ijkl})} \nonumber \\
&&\hspace{1cm}+ U_{U2}^{p-( n_{ijklm} - n_{ijkl})} + U_{UT2}^{p-(
n_{ijklm} - n_{ijkl})} + U_{UU3}^{p-( n_{ijklm} - n_{ijkl})}, \nonumber \\
&&K_{DB2} = U_{N}^{p- (n_{ijkl} + n_{ijk})} + U_{E}^{p- (n_{ijkl} + 
n_{ijk})} + U_{T}^{p- (n_{ijkl} + n_{ijk})} + U_{UB2}^{p- (n_{ijkl} + 
n_{ijk})} + U_{UD3}^{p- (n_{ijkl} + n_{ijk})} + U_{UB3}^{p- (n_{ijkl} + 
n_{ijk})} \nonumber \\
&&\hspace{1cm} + U_{U3}^{p- (n_{ijkl} + n_{ijk})}, \nonumber \\
&&K_{D2} = U_{N}^{p - n_{ijkl}} + U_{E}^{p - n_{ijkl}} + U_{T}^{p - 
n_{ijkl}} + U_{UD3}^{p - n_{ijkl}} + U_{UB3}^{p - n_{ijkl}} + U_{UT3}^{p - 
n_{ijkl}}, \nonumber \\
&&K_{DT2} = U_{N}^{p - (n_{ijkl} - n_{ijk})} + U_{E}^{p - (n_{ijkl} -
n_{ijk})} + U_{T}^{p - (n_{ijkl} - n_{ijk})} + U_{U1}^{p - (n_{ijkl} -
n_{ijk})} + U_{UT2}^{p - (n_{ijkl} - n_{ijk})} + U_{UD3}^{p - (n_{ijkl} -
n_{ijk})} \nonumber \\
&&\hspace{1cm}+ U_{U3}^{p - (n_{ijkl} - n_{ijk})} + U_{UT3}^{p - (n_{ijkl}
- n_{ijk})}, \nonumber \\
&&K_{D1} = U_{N}^{p -n_{ijk}} + U_{E}^{p -n_{ijk}} + U_{T}^{p -n_{ijk}} +
U_{UB2}^{p -n_{ijk}} + U_{UD3}^{p -n_{ijk}} + U_{UB3}^{p -n_{ijk}} +
U_{UU3}^{p -n_{ijk}}, \nonumber \\
&&K_{B} = U_{N}^{p- n_{ij}} + U_{E}^{p- n_{ij}} + U_{U1}^{p- n_{ij}} + 
U_{UB2}^{p- n_{ij}} + U_{U2}^{p- n_{ij}} + U_{UT2}^{p- n_{ij}} + 
U_{UD3}^{p- n_{ij}} + U_{UB3}^{p- n_{ij}} + U_{U3}^{p- n_{ij}} + 
U_{UT3}^{p- n_{ij}} + U_{UU3}^{p- n_{ij}}, \nonumber \\ 
&&K_{W} = U_{N}^{p-n_j} + U_{T}^{p-n_j} + U_{U1}^{p-n_j} + 
U_{UB2}^{p-n_j} + U_{U2}^{p-n_j} + U_{UT2}^{p-n_j} + U_{UD3}^{p-n_j} + 
U_{UB3}^{p-n_j} + U_{U3}^{p-n_j} + U_{UT3}^{p-n_j} + U_{UU3}^{p-n_j}, 
\nonumber \\
&&K_{S} = U_{E}^{p-1} + U_{T}^{p-1} + U_{U1}^{p-1} + U_{UB2}^{p-1} + 
U_{U2}^{p-1} + U_{UT2}^{p-1} + U_{UD3}^{p-1} + U_{UB3}^{p-1} + 
U_{U3}^{p-1} + U_{UT3}^{p-1} + U_{UU3}^{p-1}, \nonumber \\
&&K_{N} = L_{DD3}^p U_{N}^{p - (n_{ijklm} + n_{ijkl})} + L_{DB3}^p 
U_{N}^{p - (n_{ijklm} + n_{ijk})} + L_{D3}^p U_{N}^{p - n_{ijklm}} + 
L_{DT3}^p U_{N}^{p - (n_{ijklm} - n_{ijk})} \nonumber \\
&&\hspace{1cm}+ L_{DU3}^p U_{N}^{p - (n_{ijklm} - n_{ijkl})} + L_{DB2}^p 
U_{N}^{p - (n_{ijkl} + n_{ijk})} + L_{D2}^p U_{N}^{p - n_{ijkl}} + 
L_{DT2}^p U_{N}^{p - (n_{ijkl} - n_{ijk})} +L_{D1}^p U_{N}^{p - n_{ijk}} 
\nonumber \\
&&\hspace{1cm}+ L_{B}^p U_{N}^{p - n_{ij}} + L_{W}^p U_{N}^{p -n_j}, 
\nonumber \\
&&K_{E} = L_{DD3}^p U_{E}^{p - (n_{ijklm} + n_{ijkl})} + L_{DB3}^p 
U_{E}^{p - (n_{ijklm} + n_{ijk})} + L_{D3}^p U_{E}^{p - n_{ijklm}} + 
L_{DT3}^p U_{E}^{p - (n_{ijklm} - n_{ijk})} \nonumber \\
&&\hspace{1cm}+ L_{DU3}^p U_{E}^{p - (n_{ijklm} - n_{ijkl})} + L_{DB2}^p 
U_{E}^{p - (n_{ijkl} + n_{ijk})} + L_{D2}^p U_{E}^{p - n_{ijkl}} + 
L_{DT2}^p U_{E}^{p - (n_{ijkl} - n_{ijk})} +L_{D1}^p U_{E}^{p - n_{ijk}} 
\nonumber \\
&&\hspace{1cm}+ L_{B}^p U_{E}^{p - n_{ij}} + L_{S}^p U_{E}^{p - 1}, 
\nonumber \\
&&K_{T} = L_{DD3}^p U_{T}^{p - (n_{ijklm} + n_{ijkl})} + L_{DB3}^p 
U_{T}^{p - (n_{ijklm} + n_{ijk})} + L_{D3}^p U_{T}^{p - n_{ijklm}} + 
L_{DT3}^p U_{T}^{p - (n_{ijklm} - n_{ijk})} \nonumber \\
&&\hspace{1cm}+ L_{DU3}^p U_{T}^{p - (n_{ijklm} - n_{ijkl})} + L_{DB2}^p 
U_{T}^{p - (n_{ijkl} + n_{ijk})} + L_{D2}^p U_{T}^{p - n_{ijkl}} + 
L_{DT2}^p U_{T}^{p - (n_{ijkl} - n_{ijk})} + L_{D1}^p U_{T}^{p - n_{ijk}} 
\nonumber \\
&&\hspace{1cm}+ L_{W}^p U_{T}^{p - n_j} + L_{S}^p U_{T}^{p - 1}, 
\nonumber \\
&&K_{U1} = L_{DD3}^p U_{U1}^{p - (n_{ijklm} + n_{ijkl})} + L_{DT3}^p
U_{U1}^{p - (n_{ijklm} - n_{ijk})} + L_{DU3}^p U_{U1}^{p - (n_{ijklm} -
n_{ijkl})} + L_{DT2}^p U_{U1}^{p - (n_{ijkl} - n_{ijk})} \nonumber \\
&&\hspace{1cm}+ L_{B}^p U_{U1}^{p - n_{ij}} + L_{W}^p U_{U1}^{p - n_j} + 
L_{S}^p U_{U1}^{p - 1}, \nonumber \\
&&K_{UB2} = L_{DB3}^p U_{UB2}^{p - (n_{ijklm} + n_{ijk})} + L_{D3}^p
U_{UB2}^{p - n_{ijklm}} + L_{DU3}^p U_{UB2}^{p - (n_{ijklm} - n_{ijkl})} +
L_{DB2}^p U_{UB2}^{p - (n_{ijkl} + n_{ijk})} + L_{D1}^p U_{UB2}^{p -
n_{ijk}} \nonumber \\
&&\hspace{1cm}+ L_{B}^p U_{UB2}^{p - n_{ij}} + L_{W}^p U_{UB2}^{p - 
n_j} + L_{S}^p U_{UB2}^{p - 1}, \nonumber \\
&&K_{U2} = L_{DB3}^p U_{U2}^{p - (n_{ijklm} + n_{ijk})} + L_{DT3}^p 
U_{U2}^{p - (n_{ijklm} - n_{ijk})} + L_{DU3}^p U_{U2}^{p - (n_{ijklm} - 
n_{ijkl})} + L_{B}^p U_{U2}^{p - n_{ij}} + L_{W}^p U_{U2}^{p - n_j} 
\nonumber \\
&&\hspace{1cm}+ L_{S}^p U_{U2}^{p- 1}, \nonumber \\
&&K_{UT2} = L_{D3}^p U_{UT2}^{p - n_{ijklm}} + L_{DT3}^p U_{UT2}^{p - 
(n_{ijklm} - n_{ijk})} + L_{DU3}^p U_{UT2}^{p - (n_{ijklm} - n_{ijkl})} + 
L_{DT2}^p U_{UT2}^{p - (n_{ijkl} - n_{ijk})} + L_{B}^p U_{UT2}^{p - 
n_{ij}} \nonumber \\
&&\hspace{1cm} + L_{W}^p U_{UT2}^{p - n_j} + L_{S}^p U_{UT2}^{p - 1}, 
\nonumber \\
&&K_{UD3} = L_{DD3}^p U_{UD3}^{p - (n_{ijklm} + n_{ijkl})} + L_{DB2}^p
U_{UD3}^{p - (n_{ijkl} + n_{ijk})} + L_{D2}^p U_{UD3}^{p - n_{ijkl}} +
L_{DT2}^p U_{UD3}^{p - (n_{ijkl} - n_{ijk})} + L_{D1}^p U_{UD3}^{p -
n_{ijk}} \nonumber \\
&&\hspace{1cm}+ L_{B}^p U_{UD3}^{p - n_{ij}} + L_{W}^p U_{UD3}^{p - 
n_j} + L_{S}^p U_{UD3}^{p - 1}, \nonumber \\
&&K_{UB3} = L_{DB3}^p U_{UB3}^{p - (n_{ijklm} + n_{ijk})} + L_{DB2}^p 
U_{UB3}^{p - (n_{ijkl} + n_{ijk})} + L_{D2}^p U_{UB3}^{p - n_{ijkl}} + 
L_{D1}^p U_{UB3}^{p - n_{ijk}} + L_{B}^p U_{UB3}^{p - n_{ij}} + L_{W}^p 
U_{UB3}^{p - n_j} \nonumber \\
&&\hspace{1cm}+ L_{S}^p U_{UB3}^{p - 1}, \nonumber \\
&&K_{U3} = L_{DB2}^p U_{U3}^{p - (n_{ijkl} + n_{ijk})} + L_{DT2}^p 
U_{U3}^{p - (n_{ijkl} - n_{ijk})} + L_{B}^p U_{U3}^{p - n_{ij}} + L_{W}^p 
U_{U3}^{p - n_j} + L_{S}^p U_{U3}^{p - 1}, \nonumber \\
&&K_{UT3} = L_{DT3}^p U_{UT3}^{p - (n_{ijklm} - n_{ijk})} + L_{D2}^p
U_{UT3}^{p - n_{ijkl}} + L_{DT2}^p U_{UT3}^{p - (n_{ijkl} - n_{ijk})} +
L_{B}^p U_{UT3}^{p - n_{ij}} + L_{W}^p U_{UT3}^{p - n_j} + L_{S}^p
U_{UT3}^{p - 1}, \nonumber \\
&&K_{UU3} = L_{DU3}^p U_{UU3}^{p - (n_{ijklm} - n_{ijkl})} + L_{D1}^p
U_{UU3}^{p - n_{ijk}} + L_{B}^p U_{UU3}^{p - n_{ij}} + L_{W}^p U_{UU3}^{p
- n_j} + L_{S}^p U_{UU3}^{p - 1},
\end{eqnarray}

\end{document}